\documentclass{article}
\usepackage[doublespacing]{setspace}

\usepackage{amsmath}
\usepackage{amsfonts}
\usepackage{graphicx}
\usepackage{geometry}

\usepackage[tight]{subfigure}
\newcommand{\avg}[1]{\left< #1 \right>} 

\begin{document}


\title{Modular Random Boolean Networks\footnote{A preliminary version of this work was presented at the ALife XII conference in Odense, Denmark on August $20^{th}$, 2010. \cite{BalpoGershenson:2010}}}

\author{
Rodrigo Poblanno-Balp$\dagger\ddagger$\\
Carlos Gershenson$\ddagger\dagger$\\
\small $\dagger$ Centro de Ciencias de la Complejidad, UNAM, M\'exico\\
\small $\ddagger$ Instituto de Investigaciones en Matem\'aticas Aplicadas y en Sistemas, UNAM, M\'exico}

\maketitle

\begin{abstract}
Random Boolean networks (RBNs) have been a popular model of genetic regulatory networks for more than four decades. However, most RBN studies have been made with random topologies, while real regulatory networks have been found to be modular. 
In this work, we extend classical RBNs to define modular RBNs.
Statistical experiments and analytical results show that modularity has a strong effect on the properties of RBNs. In particular, modular RBNs have more attractors and are closer to criticality when chaotic dynamics would be expected, compared to classical RBNs.
\end{abstract}

\section{Introduction}

Random Boolean networks (RBN) have been a popular model of genetic regulatory networks (GRNs) \cite{Kauffman1969,Kauffman1993,Gershenson2004c}. Most studies have been made on RBNs with random topologies. Nevertheless, it has been found that topologies affect considerably the properties of RBNs. For example, Aldana studied RBNs with a scale-free topology \cite{Aldana2003}, discovering important differences with random topologies. In this work, we study the effect of a modular topology in RBNs. We find that modularity changes the properties of RBNs. Given the fact that real GRNs are modular \cite{segal,callebaut05,schlosser} and most RBN studies have been made over random topologies, it is important to understand the differences between random and modular topologies. 

Modularity plays an important role in evolution \cite{Simon1996,FernandezSole2003,Watson:2006}, since separable functional systems are found at all scales of biological systems \cite{wagner07}.
Modularity allows for changes to occur within modules without propagating to other regions and the combination of modules to explore new functions \cite{espinoza-soto}. Thus, the study of modular RBNs is also relevant for understanding the evolution of GRNs. 

In the next section, classic RBNs are reviewed, together with their dynamical properties and related work. Section \ref{sec:mrbns} presents our model of modular RBNs. Methods and results of statistical experiments follow in Section \ref{sec:experiments}. The discussion in Section \ref{sec:discussion} reflects on the results and provides an analytical confirmation. Several future research avenues are mentioned to conclude the paper. 

\section{Random Boolean Networks}

\begin{figure}[htbp]
\begin{center}
\includegraphics[width=.5\textwidth]{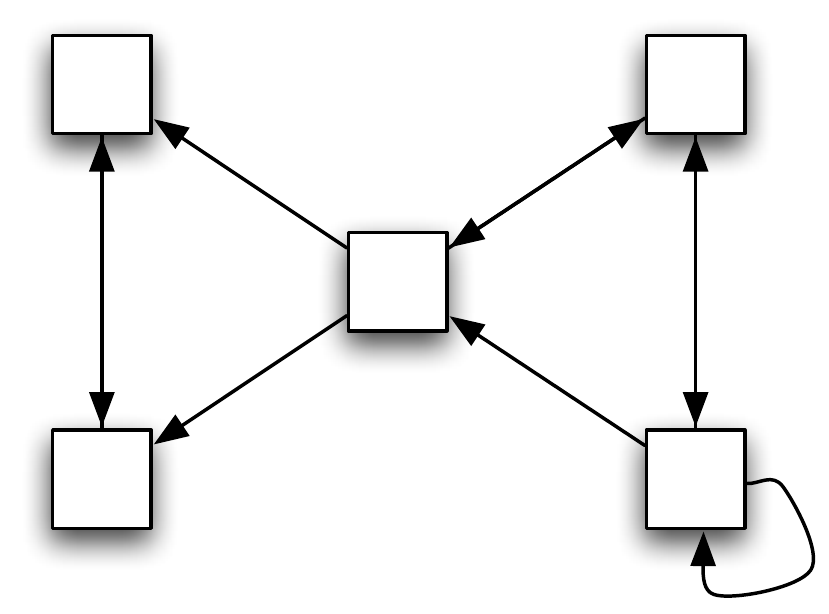}
\caption{Topology of an example RBN with $N=5$, $K=2$. Each node has two inputs that determine its state. Since the topology is randomly generated, a node might have several outputs or none at all.}
\label{fig:topoRBN}
\end{center}
\end{figure}

Random Boolean Networks (RBNs) \cite{Kauffman1969,Kauffman1993,Gershenson2004c} consist of $N$ nodes with a Boolean state, representing whether a gene is active (``on'' or ``one'') or inactive (``off'' or ``zero''). These states are determined by the states of $K$ nodes which can be considered as inputs or links towards a node. Because of this, RBNs are also known as NK networks or Kauffman models \cite{AldanaEtAl2003}. The states of nodes are decided by lookup tables that specify for every $2^K$ possible combination of input states the future state of the node. RBNs are random in the sense that the connectivity (which nodes are inputs of which, see Figure \ref{fig:topoRBN}) and functionality (lookup tables of each node, see Table \ref{tab:rules}) are chosen randomly when a network is generated, although these remain fixed as the network is updated each time step. RBNs are discrete dynamical networks (DDNs), since they have discrete values, number of states, and time \cite{Wuensche1998}. They can also be seen as a generalization of Boolean cellular automata \cite{Wolfram1986,Gershenson2002e}, where each node has a different neighborhood and rule.

\begin{table}[htbp]
\caption{Example lookup table to update a node $z$ depending on the state of nodes $x$ and $y$. Lookup tables include all possible combinations of inputs, i.e. $2^K$ rows. Different nodes will have different lookup tables, i.e. Boolean functions. In this example table, the function is the binary XNOR.}
\label{tab:rules} 
\begin{center}
\begin{tabular}{lll}
\hline\noalign{\smallskip}
x(t) & y(t) & z(t+1)  \\
\noalign{\smallskip}\hline\noalign{\smallskip}
0 & 0 & 1 \\
0 & 1 & 0 \\
1 & 0 & 0 \\
1 & 1 & 1 \\
\noalign{\smallskip}\hline
\end{tabular}
\end{center}
\end{table}

RBNs have $2^N$ possible network states, i.e. all possible combinations of Boolean node states. Transitions between network states determine the state space of the RBN. In classic RBNs, the updating is deterministic and synchronous \cite{Gershenson2002e}. Since the number of states of the network is finite and the dynamics are deterministic, sooner or later a state will be repeated in theory (in practice, this can take longer than the age of the universe due to the immense state space). When this occurs, the network has reached an $attractor$, since the dynamics will remain in that subset of the state space. If the attractor consists of only one state, then it is called a \emph{point} attractor (similar to a steady state), whereas an attractor consisting of several states is called a \emph{cycle} attractor (similar to a limit cycle).

\begin{figure}[htbp]
\begin{center}
\includegraphics{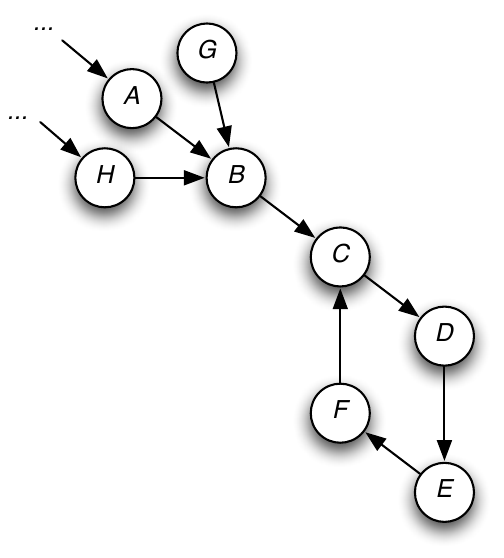}
\caption{Example of state transitions. $B$ is a successor state of $A$ and a predecessor of $C$. States can have many predecessors (e.g. $B$), but only one successor. $G$ is a Garden of Eden state since it has no predecessors. The attractor $C\rightarrow D\rightarrow E \rightarrow F \rightarrow C$ has a period four. \cite{Gershenson:2010}}
\label{fig:statesRBN}
\end{center}
\end{figure}

RBNs are dissipative systems, since each state has only one successor, while having the possibility of having several predecessor states (many states lead to one state), or no predecessor (a state can be reached only from initial conditions, a so called a ``Garden of Eden" state). Figure \ref{fig:statesRBN} illustrates the state transitions of RBNs.

Note that the topological network (with $N$ nodes, each with a Boolean variable, i.e. one bit) is different from the state transition network (with $2^N$ nodes, each with $N$ bits). One of the main avenues in RBN research is involved with studying how the topological network (structure) determines the properties of the state transition network (function).

\subsection{Dynamical Regimes}

\begin{figure}
     \centering
     \subfigure[]{
          \label{fig:dynA}
          \includegraphics[width=.9\textwidth]{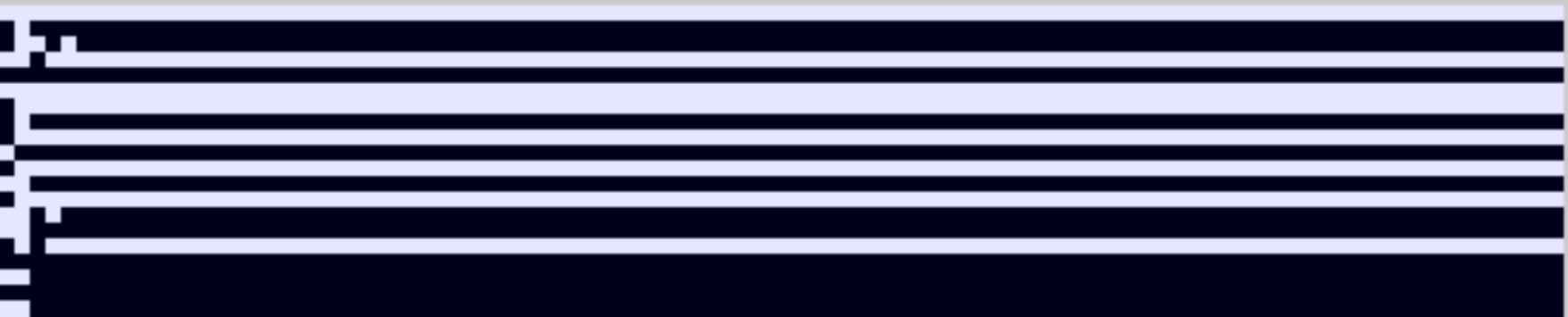}
	}\\
     \subfigure[]{
          \label{fig:dynB}
          \includegraphics[width=.9\textwidth]{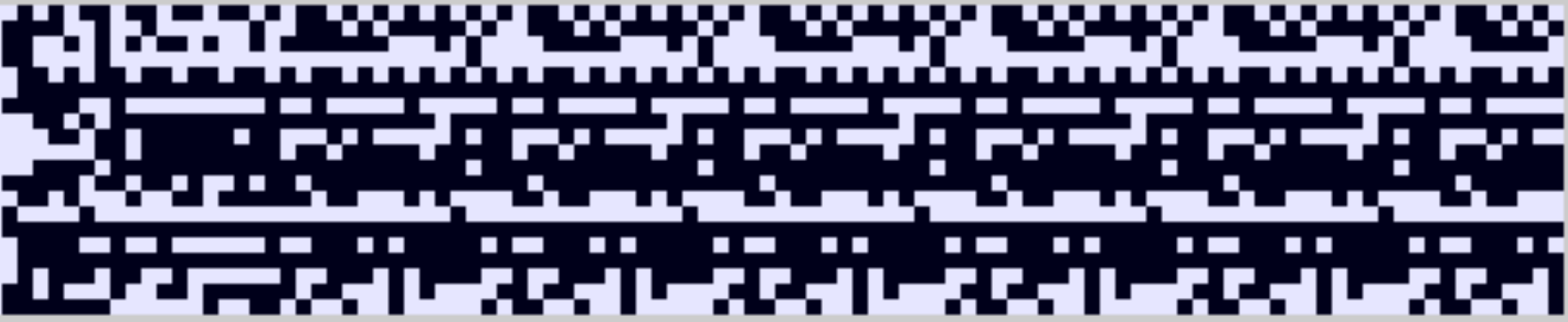}
	}\\
     \subfigure[]{
          \label{fig:dynC}
          \includegraphics[width=.9\textwidth]{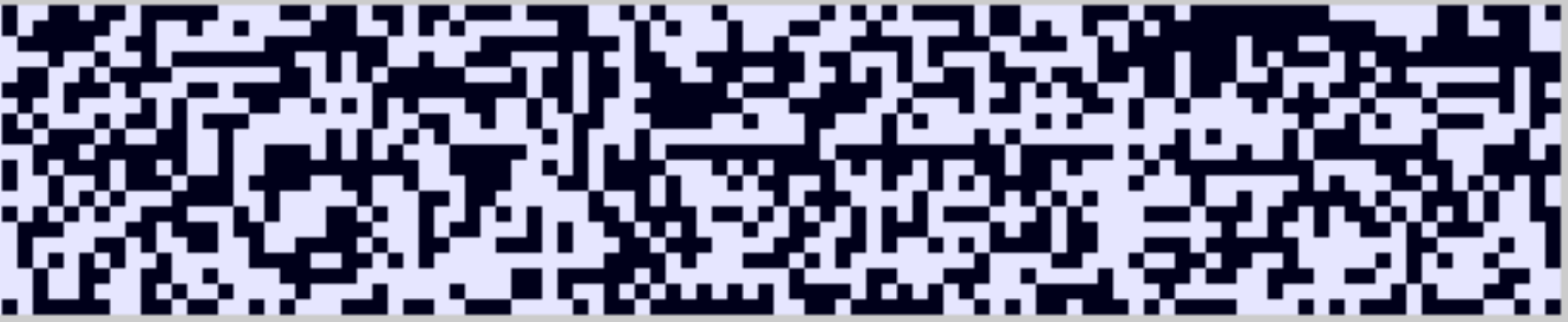}
	}

     \caption{Dynamics for three RBN with $N=20$ and $p=0.5$: (A) $K=1$ (ordered), (B) $K=2$ (critical), and (C) $K=5$ (chaotic). 100 time steps shown from a random initial condition (leftmost column, time flows to the right, the state of nodes in time is represented in rows, the RBN state at a particular timestep is represented in columns).}
     \label{fig:dyn}
\end{figure}

RBNs have three dynamical regimes: \emph{ordered}, \emph{chaotic}, and \emph{critical} \cite{Wuensche1998,Gershenson2004c}. Typical dynamics of the three regimes can be seen in Figure \ref{fig:dyn}. The ordered regime is characterized by little change, i.e. most nodes are static. The chaotic regime is characterized by large changes, i.e. most nodes are changing. This implies that RBNs in the ordered regime are robust to perturbations (of states, of connectivity, of node functionality). Since most nodes do not change, damage has a low probability of spreading through the network. On the contrary, RBNs in the chaotic regime are very fragile: since most nodes are changing, damage spreads easily, creating large avalanches that spread through the network. The critical regime balances the ordered and chaotic properties: the network is robust to damage, but it it not static. This balance has led people to argue that life and computation should be within or near the critical regime \cite{Langton1990,Kauffman1993,Crutchfield:1994,Kauffman2000}. In the ordered regime, there is robustness, but no possibility for dynamics, computation, and exploration of new configurations, i.e. evolution. In the chaotic regime, exploration is possible, but the configurations found are fragile, i.e. it is difficult to reach persisting patterns (memory). There is recent evidence that real GRNs are in or near the critical regime \cite{Balleza:2008}.

It has been found that the regimes of RBNs depend on several parameters and properties \cite{Gershenson:2010}. Still, two of the most salient parameters are the connectivity $K$ and the probability $p$ that there is a one on the last column of lookup tables. When $p=0.5$ there is no probability bias. For $p=0.5$, the ordered regime is found when $K<2$, the chaotic regime when $K>2$, and the critical regime when $K=2$ \cite{DerridaPomeau1986}. The ordered and chaotic regimes are found in distinct phases, while the critical regime is found on the phase transition. Derrida and Pomeau found analytically the critical connectivity $K_c$\footnote{This result is for infinite-sized networks. In practice, for finite-sized networks, the precise criticality point may be slightly shifted.}:

\begin{equation}
\avg{K_c} = \frac{1}{2p(1-p)}
\label{eq:kc}
\end{equation}

This can be explained using the simple method of Luque and Sol\'{e} \cite{LuqueSole1997}: Focussing on a single node $i$, the probability that a damage to it will percolate through the network can be calculated. It can be seen that this probability will increase with $K$, as more outputs from a node will increase the chances of damage propagation. Focussing on a node $j$ from the outputs of $i$, then there will be a probability $p$ that $j=1$. Thus, there will be a probability $1-p$ that a damage in $i$ will propagate to $j$. Complementarily, there will be a probability $1-p$ that $j=0$, with a probability $p$ that a damage in $i$ will propagate to $j$. If there are on average $\avg{K}$ nodes that $i$ can affect, then 
we can expect damage to spread if $\avg{K}2p(1-p)\geq 1$ \cite{LuqueSole1997}, which implies chaotic dynamics. This leads to the critical connectivity of Derrida and Pomeau \cite{DerridaPomeau1986}, shown in equation \ref{eq:kc}.

\subsection{Related Work}

We briefly mention particular studies of coupled RBNs, which share similarities with the model of modular RBNs presented in the next section. A more detailed comparison can be found elsewhere \cite{Balpo:2011}.

Bastolla and Parisi \cite{Bastolla:1998} studied modularity within classical RBNs, i.e. functionally independent clusters, but not topological modularity.

There have been different studies where only two coupled RBNs are considered \cite{ho05,andrecut05,hung06}. 

There are studies where RBNs are generated in cells of a 2D lattice, similar to a cellular automaton, where each RBN is weakly coupled with its von Neumann neighbors \cite{villani06,serra08,Damiani:2010}. The goal is to model intercellular signaling in a tissue.

Our model is more general than previous models, since there is an arbitrary number of coupled networks, and this is not restricted to spatial neighbors. Moreover, it is a natural extension of the classic RBN model.

\section{Modular Random Boolean Networks}
\label{sec:mrbns}

\begin{figure}[htbp]
\begin{center}
\includegraphics[width=.75\textwidth]{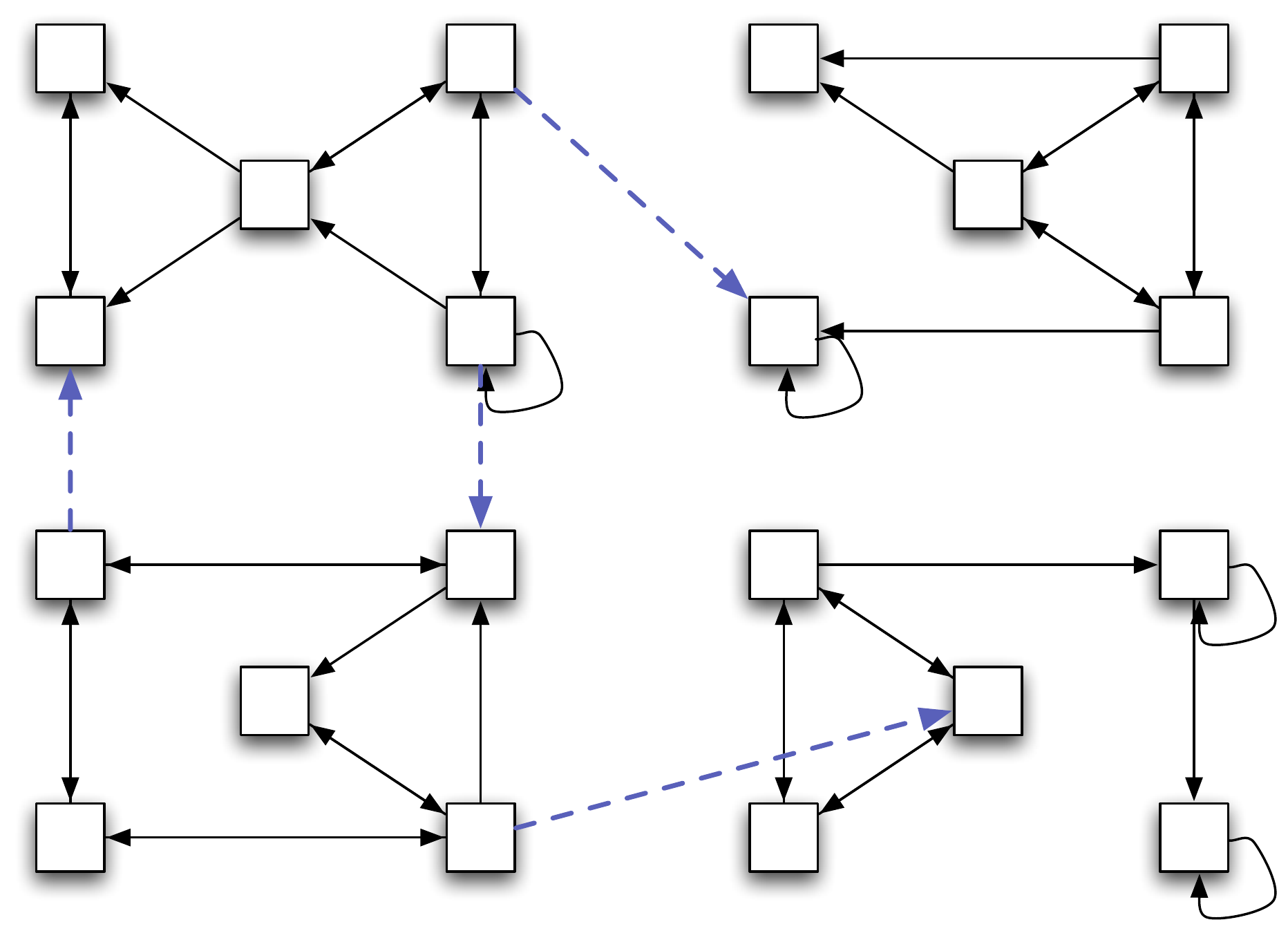}
\caption{Topology of a example MRBN with $N=5$, $M=4$, $K=2$, $L=1$. Each module is similar to the example RBN shown in Figure \ref{fig:topoRBN}, with one additional input per module (dashed arrows). Modules can have several or no outputs.}
\label{fig:topoMRBN}
\end{center}
\end{figure}

We propose a general model of modular random Boolean networks (MRBNs) that extend naturally the classic RBN model. A MRBN consists of $M$ modules, each of which is a RBN with $N$ nodes and on average $\avg{K}$ (intramodular) inputs per node. Each module has additional $\avg{L}$ (intermodular) inputs on average that link random nodes from different modules. These $L$ intermodular connections can be seen as ``weak links'' \cite{Csermely:2006} between modules. Weak links have been shown to offer stability in networks \cite{Csermely:2006}. Figure \ref{fig:topoMRBN} shows an example MRBN.

Thus, the total number of nodes $N_T$ of a MRBN is given by
\begin{equation}
N_T = N\cdot M
\label{eq:nt}
\end{equation}
and the total number of links $T$ is given by 
\begin{equation}
T = M\cdot (\avg{K}\cdot N + \avg{L})
\label{eq:t}
\end{equation}
since each of the $M$ modules has $\avg{L}$ inputs and  $N$ nodes with $\avg{K}$ intramodular inputs.

The total average number of inputs per node $\avg{K_T}$ is given by 
\begin{equation}
\avg{K_T} =\dfrac{T}{N_T} = \avg{K} + \dfrac{\avg{L}}{N}
\label{eq:kt}
\end{equation}

In the exploration of the space of possible MRBNs, the following measures are useful:

To study the relationship between number of nodes and modules, the node-to-module ratio $\mu$ is simply
\begin{equation}
\mu = \dfrac{N}{M}
\label{eq:miu}
\end{equation}

To study the relationship between internal ($K$) and external ($L$) links, the probability $\kappa$ that a link is \emph{intramodular} is given by 
\begin{equation}
 \kappa = \dfrac{\avg{K}}{\avg{K_T}}
\label{eq:kappa}
\end{equation}
while the probability $\lambda$ that a link is \emph{intermodular} is the complement of $\kappa$:
\begin{equation}
 \lambda= 1 - \kappa
\label{eq:lambda}
\end{equation}

\section{Experiments}
\label{sec:experiments}

The open software laboratory RBNLab \cite{RBNLab} was extended to explore the properties of MRBNs. RBNLab and its Java source code are available at http://rbn.sourceforge.net.

For all experiments, $p=0.5$ and a total number of nodes $N_T=20$ was used. Even when this is a relatively small size of MRBN, the effects of modularity can be already appreciated. The size of networks severely limits the statistical explorations, since each additional node doubles the size of the state space $S=2^{N_T}$.

For each case and each $K_T$, one thousand networks were generated randomly, exploring one thousand randomly chosen initial states for ten thousand steps. This implies at least $10^{10}$ updates per MRBN ensemble \cite{Kauffman2004}.

We performed two sets of experiments: one to explore the statistical properties of different families of MRBNs and another to measure the sensitivity to initial conditions of different families of MRBNs. In each set of experiments, five cases were studied in two groups. In the first group, $\kappa$ (percentage of internal inputs) is explored, while leaving $\mu$ (node-to-module ratio) fixed. In the second group, $\mu$ is varied while $\avg{K} = \avg{L}$. 
For both sets of experiments, the following five cases were considered, varying $\avg{K_T} = \{1,2,3,4\}$:

\begin{enumerate}
 \item $\avg{K} = \avg{L}$,\quad \quad $\mu\rightarrow1$
 \item $\avg{K} = 1$,\quad\quad$\mu\rightarrow1$
 \item $\avg{L} = 1$,\quad\quad$\mu\rightarrow1$
 \item $N = 20$,\quad $M = 1$, \quad $\avg{L} = 0$\quad, \quad $\avg{K}=\avg{K_T}$
 \item $N = 1$, \quad$M = 20$, \quad $\avg{L} = \avg{K}$
\end{enumerate}
 
For cases 1, 2 and 3, $\mu$ remains fixed ($\mu=\frac{N}{M}\rightarrow1$) while $\kappa$ is explored. Case 2 favors intermodular (external) links, while keeping intramodular links fixed at $\avg{K} = 1$. Case 3 favors intramodular (internal) links, while keeping intermodular links fixed at $\avg{L} = 1$. Case 1 is intermediate, balancing intermodular and intramodular links, restricting $\avg{K} = \avg{L}$.

Cases 1, 4 and 5 are compared to explore the effect of $\mu$ on MRBN properties. Case 4 is equivalent to the classical RBN model, since there is only one module ($M = 1$) and no intermodular links ($\avg{L} = 0$). Case 5 is the opposite extreme, where there are $N_T$ modules of a single node. In this case, intramodular links are self-links, and when $\avg{K} > 1$ there are in practice fictitious inputs, since the dynamic behavior is equivalent to that of $\avg{K} = 1$. This is because having more than one input from the same node is equivalent to having only one input (zero or one), as the same node cannot have different states at the same time. Case 5 is different from classical RBNs with only self-links since $\avg{L} > 0$.

In the following, the variables representing averages such as  $\avg{K}$ and $\avg{L}$ will be used without the average symbols for simplicity.

\subsection{Statistical Properties}

For the statistical experiments, the following properties were studied:

\begin{description}
\item[Average Number of Attractors $A$.]  This indicates how many distinct sets of states can ``attract" the dynamics of the MRBN.  When $A>1$ it is considered that the system is \emph{multistable} \cite{Thomas1978}. There is evidence that in real genetic regulatory networks, attractors correspond to cell types \cite{HuangIngber2000}, confirming Kauffman's original hypothesis \cite{Kauffman1969}.
\item[Average Attractor Lengths $Le$.] When $Le=1$, there are only point attractors in the network. Larger values of $Le$ indicate longer cycle attractors. 
\item[Average Percentage of States in Attractors $\%SIA$.] This reflects how much ``complexity reduction" is performed by the network, i.e. from all possible $2^{N_T}$ states, the percentage of states that ``capture" the dynamics. Even when complexity reduction is relevant, larger values of $\%SIA$ indicate a more complex potential functionality of the network, i.e. richer dynamics. A large $\%SIA$ can be given by large attractor lengths $Le$ and/or a high number of attractors $A$. The more and longer attractors a network has, it can exhibit a richer behavior. 
\end{description}

\begin{equation}
\%SIA=100\cdot \dfrac{A\cdot Le}{2^{N_T}}
\label{eq:sia}
\end{equation}

Statistical results often give very high standard deviations $\sigma$. This is because some networks might have a single point attractor, while others might have several cycle attractors. Still, the averaged values are informative, showing  the effect of different MRBN parameters on the network properties and dynamics. Nevertheless, the potential implications of very high standard deviations should not be forgotten. For example, in classic RBNs $K=N$ with $p=0.5$ implies chaotic dynamics. Still, it is possible to construct within this ensemble RBNs with ordered dynamics, e.g. having a large number of canalizing functions \cite{Harris:2002,Shmulevich:2004,Kauffman:2004}.

\subsubsection{Number of Attractors}

\begin{figure}[ht!]
\begin{center}
\includegraphics[width=0.8\textwidth]{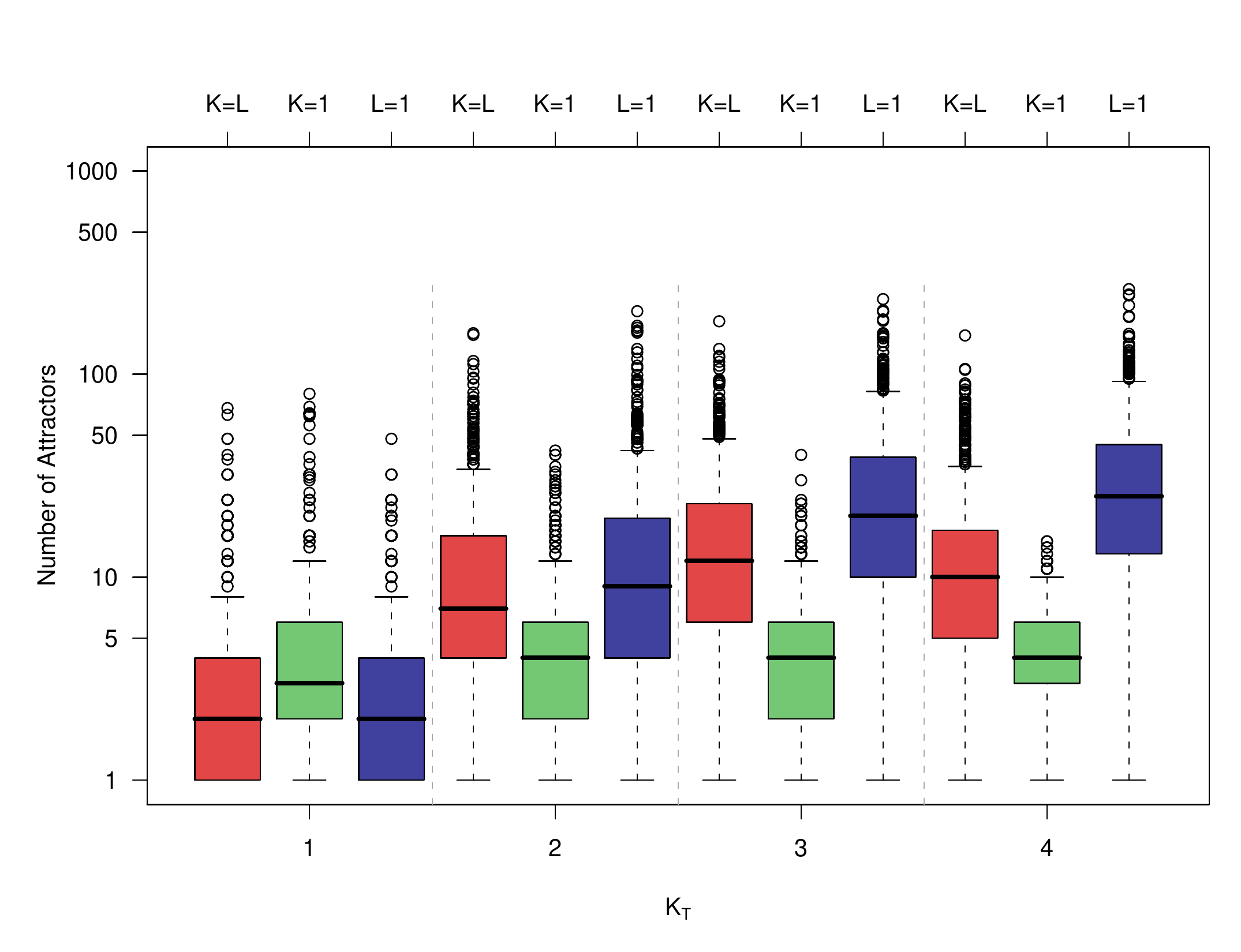}
\end{center}
 \caption{Number of Attractors for different $K_T$\ for cases 1, 2 \& 3 ($K=L$, $K=1$, and $L=1$, respectively). Notice logarithmic scale.}
\label{fig:attractors1}
\end{figure}

The $A$ results for cases 1, 2, and 3 are shown as boxplots\footnote{A boxplot is a non-parametric representation of a statistical distribution. Each box contains the following information: The median ($Q2=x_{0.50}$) is represented by the horizontal line inside the box. The lower edge of the box represents the lower quartile ($Q1=x_{0.25}$) and the upper edge represents the upper quartile ($Q3=x_{0.75}$). The interquartile range ($IQR=x_{0.75}-x_{0.25}$) is represented by the height of the box. Data which is less than $Q1 - 1.5\cdot IQR$ or greater than $Q3 + 1.5\cdot IQR$ is considered an ``outlier", and is indicated with circles. The ``whiskers" (horizontal lines connected to the box) show the smallest and largest values that are not outliers. } in Figure \ref{fig:attractors1}. The primary result is that higher values of $\kappa$ yield more attractors. Detailed results (including means and standard deviations) and $\kappa$ values can be seen in Tables \ref{tab:statsc1}, \ref{tab:statsc2}, and \ref{tab:statsc3} in Appendix \ref{sec:tables}. Notice that for $K_T=1$, case 2 has the highest $\kappa=1$, since the restriction $K=1$ implies $L=0$. For $K_T \geq 2$, case 2 has the lowest $\kappa$, and also the lowest $A$. Case 3 has the highest $\kappa$ for $K_T \geq 2$, and also the highest $A$.

These results suggest that favoring intramodular ($K$) over intermodular ($L$) links results in a higher number of attractors. This can be explained as follows: if there is a maximum $\kappa=1 \Rightarrow L=0$, i.e. there are no intermodular links. This means that modules are isolated and can be seen as independent classic RBNs, with different attractors of different lengths. However, the MRBN will consider different combinations of the same modular attractors as different attractors. This can be better understood with an example. Let us have a small MRBN $M=2, N=3, K=2, L=0$. Since modules have no interaction ($L = 0$), this MRBN is equivalent to two separate classical RBNs. Let us assume that the first module has a point attractor: $001 \rightarrow 001$ and an attractor of period 2: $000\rightarrow 111\rightarrow 000$; and the second module has a point attractor: $000 \rightarrow 000$ and an attractor of period 3: $100 \rightarrow 010 \rightarrow 001 \rightarrow 100$. Thus, the combinations of these RBN attractors will yield four attractors in the MRBN: 

\begin{enumerate}
\item The two point attractors: $001000\rightarrow 001000$.
\item The first point attractor and the period three attractor: $001100\rightarrow 001010\rightarrow 001001 \rightarrow 001100$.
\item The period two attractor and the second point attractor: $000000\rightarrow 111000\rightarrow 000000$. 
\item The period two and period three attractors: $000100\rightarrow 111010\rightarrow 000001 \rightarrow 111100\rightarrow 000010\rightarrow 111001 \rightarrow 000100$.
\end{enumerate}

Considering that in RBNs $A$ grows algebraically with $N$ \cite{Gershenson2004c}, MRBNs with several modules $M$ and $\kappa=1$ will tend to have much more attractors on average than a RBN with the same $N_T$ and $K_T$. This is because the MRBN will have as different attractors all the possible combinations of all modular attractors. As intermodular links $L$ are added and $\kappa$ decreases, changes in the states of nodes which have $L$ links as outputs might perturb and even destroy attractors. When $\kappa$ is minimal, the organization of the MRBN is more similar to a classical RBN, since there are less restrictions on where to assign links (in general (for $M>2$), there are more possible intermodular links ($M \cdot (M-1) \cdot N^2$) than possible intramodular links ($M \cdot N^2$)).

\begin{figure}[ht!]
\begin{center}
\includegraphics[width=0.8\textwidth]{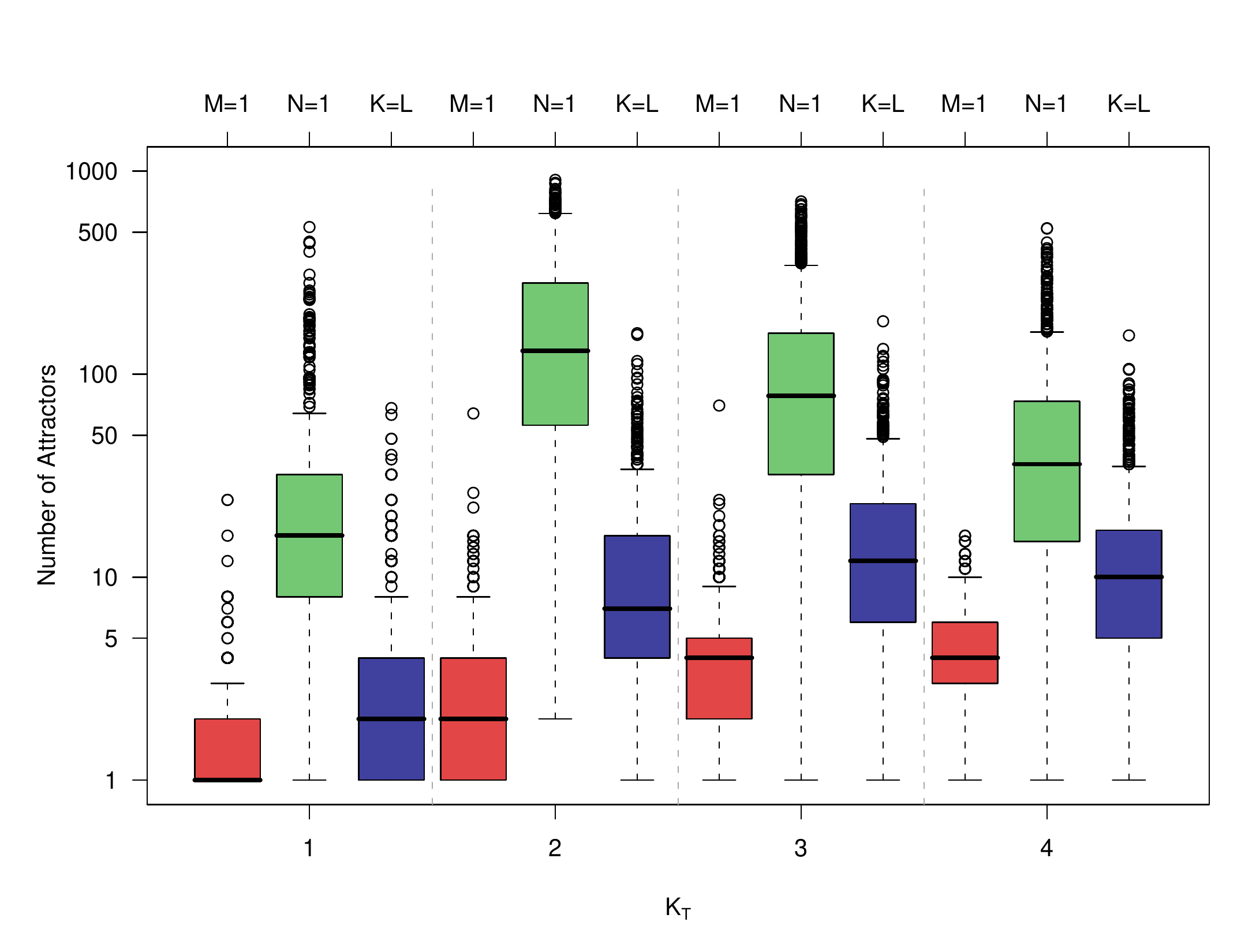}
\end{center}
 \caption{Number of Attractors for different $K_T$\ for cases 4, 5 \& 1 ($M=1$, $N=1$, and $K=L$, respectively). Notice logarithmic scale.}
\label{fig:attractors2}
\end{figure}

The $A$ results for cases 4, 5, and 1 are compared in Figure \ref{fig:attractors2}. It can be seen that $\mu$ is also relevant for $A$. More modules (low $\mu$) imply more potential combinations of modular attractors, which implies a higher $A$ for MRBNs. Case 4, which is equivalent to a classic RBN has a maximum $\mu=N_T$, i.e. a single module, so there is no possible combination of attractors. Case 4 has the lowest $A$ of all five cases. The extreme case 5 has a minimum $\mu=\dfrac{1}{N_T}$, i.e. $N=1$, giving the highest $A$ of all five cases. Notice that in case 5 $A$ decreases for $K_T>2$. This is because even when $\kappa$ is constant in theory, in practice $\kappa$ decreases as $K_T$ increases, given the fact that if $K>N$ is equivalent in practice to $K=N$ because of fictitious inputs ($N=1$ in this case, see Table \ref{tab:statsc5} in Appendix \ref{sec:tables}).

\subsubsection{Attractor Lengths}

\begin{figure}[ht!]
\begin{center}
\includegraphics[width=0.8\textwidth]{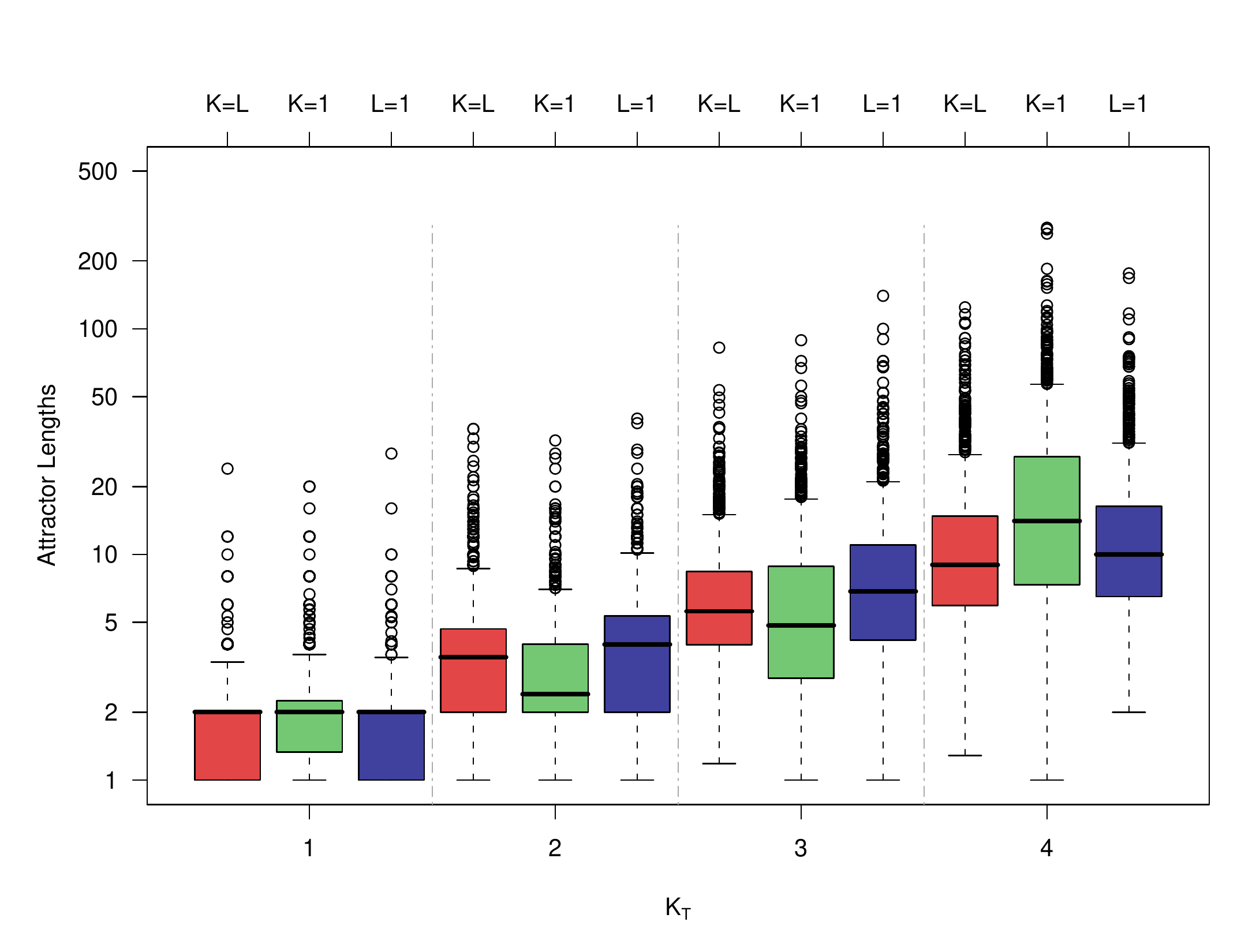}
\end{center}
 \caption{Attractor Lengths for cases 1, 2 \& 3 ($K=L$, $K=1$, and $L=1$, respectively). Notice logarithmic scale.}
\label{fig:lengths1}
\end{figure}

The effect of $\kappa$ on $Le$ seems to be minimal, as the average attractor lengths is very similar for cases 1, 2, and 3, exponentially increasing with $K_T$, as it can be seen in Figure \ref{fig:lengths1}.

\begin{figure}[ht!]
\begin{center}
\includegraphics[width=0.8\textwidth]{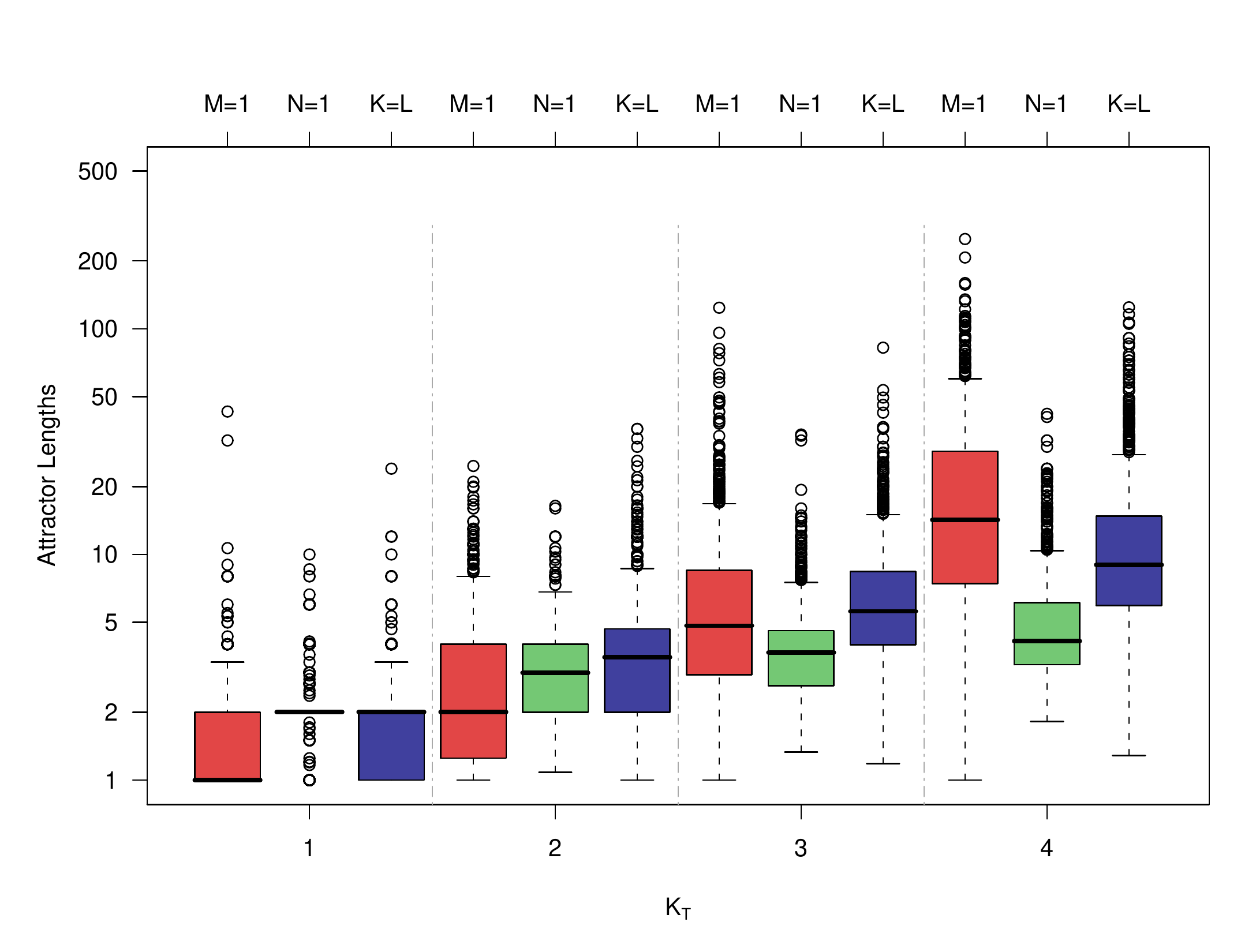}
\end{center}
 \caption{Attractor Lengths for cases 4, 5 \& 1 ($M=1$, $N=1$, and $K=L$, respectively). Notice logarithmic scale.}
\label{fig:lengths2}
\end{figure}

The effect of $\mu$ on $Le$ is seen in Figure \ref{fig:lengths2}. For $K\geq3$, classical RBNs ($\mu=1$) have a higher $Le$ than a balanced MRBN (case 1), which have a higher $Le$ than extreme MRBNs with a minimal $\mu=\dfrac{1}{N_T}$ (case 5). This can be explained by the fact that in practice attractor lengths grow algebraically with $N$ (for deterministic updating schemes, as the one used in this paper) \cite{Gershenson2004b}. For the same value of $N_T$, higher values of $\mu$ imply a higher $N$ per module. Having more nodes per module allows the possibility of more combinations of states in an attractor, increasing its length. It is not so much that a large $N$ favors longer attractors, but a small $N$ restricts their possibility.

For $K\leq2$, classical RBNs have the lowest $Le$. Here the modular cases can have longer attractors because of the combination of modular attractors is possible with a low $L$, as explained for $A$. 

\subsubsection{\% of States in Attractors}

Since all cases have a constant $N_T=20$, the $\%SIA$ depends only on $A$ and $Le$, as shown by equation \ref{eq:sia}. 

\begin{figure}[ht!]
\begin{center}
\includegraphics[width=0.8\textwidth]{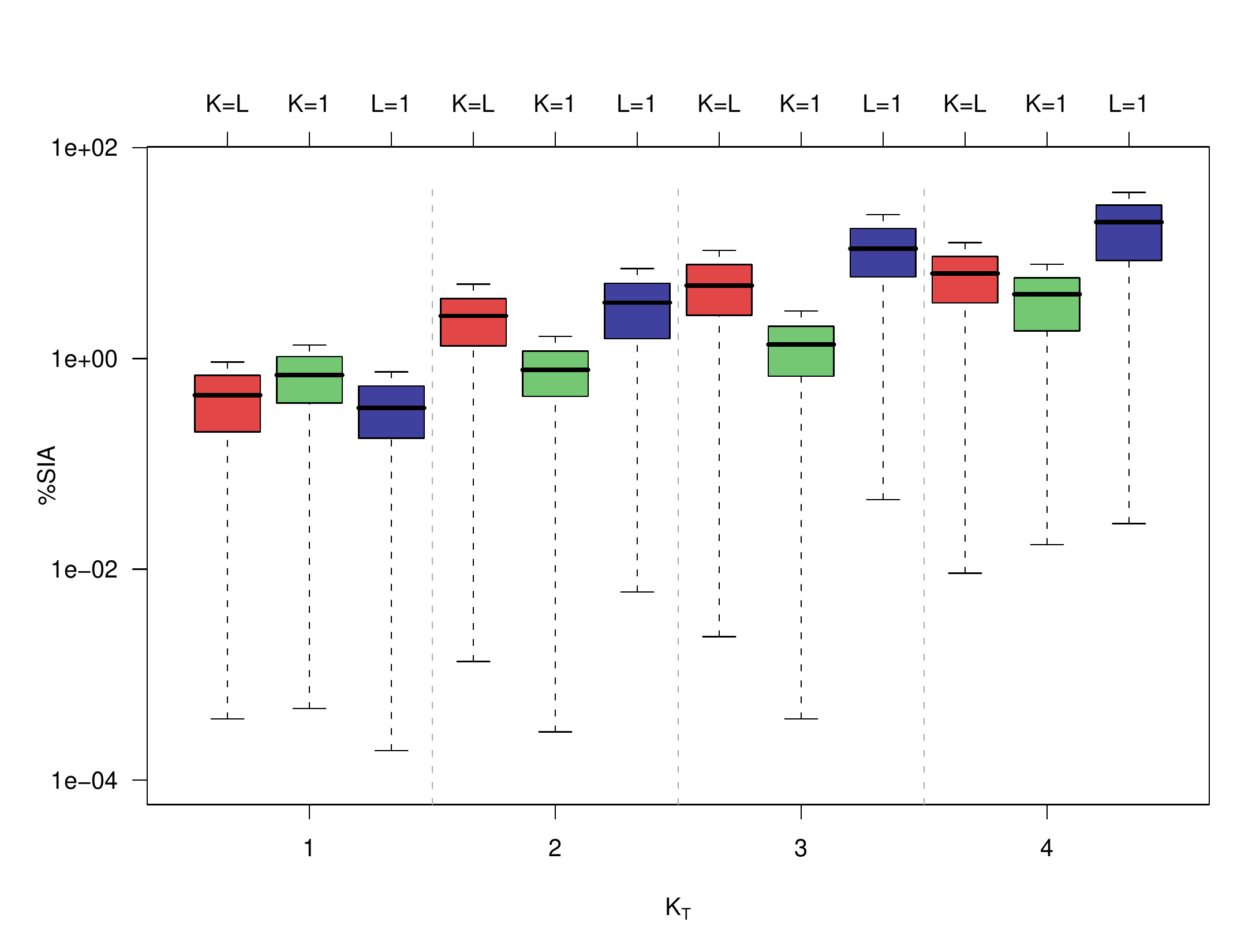}
\end{center}
 \caption{Percentage of States in Attractors for cases 1, 2 \& 3 ($K=L$, $K=1$, and $L=1$, respectively). Notice logarithmic scale.}
\label{fig:sia1}
\end{figure}

For cases where $\kappa$ was varied (1, 2, and 3), it was shown that $Le$ did not vary much depending on $\kappa$. Thus, the results for $\%SIA$ shown in Figure \ref{fig:sia1} are very similar to those of $A$ in Figure \ref{fig:attractors1}: a higher $\kappa$ yields a higher $\%SIA$ (also seen in the means shown in Tables  \ref{tab:statsc1}, \ref{tab:statsc2}, and \ref{tab:statsc3} in Appendix \ref{sec:tables}).

\begin{figure}[ht!]
\begin{center}
\includegraphics[width=0.8\textwidth]{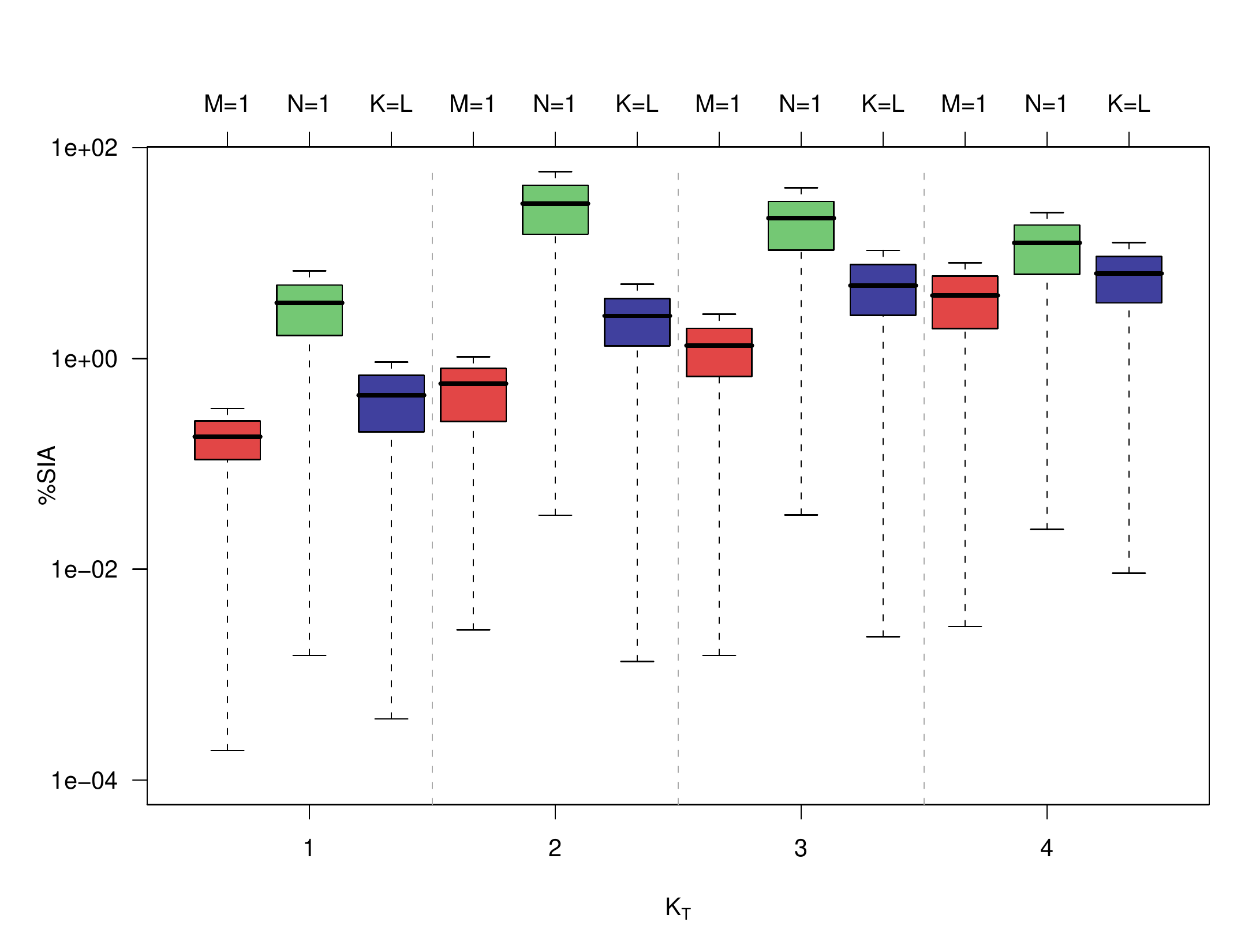}
\end{center}
 \caption{Percentage of States in Attractors for cases 4, 5 \& 1 ($M=1$, $N=1$, and $K=L$, respectively). Notice logarithmic scale.}
\label{fig:sia2}
\end{figure}

For cases where $\mu$ was varied (4, 5, and 1), shown in Figure \ref{fig:sia2}, the results of $\%SIA$ also resemble those of $A$. This is because the differences in the number of attractors (Figure \ref{fig:attractors2}) are greater than those of attractor lengths (Figure \ref{fig:lengths2}).

\subsection{Sensitivity to Initial Conditions}

One way to characterize the dynamical regime of an ensemble of discrete dynamical systems such as MRBNs is by measuring the sensitivity to initial conditions. This is done by measuring how small differences in initial states lead to similar or different states: if the system is sensitive to small differences, it is considered chaotic. 
This method is similar to damage spreading \cite{Stauffer:1994} or stability analysis of dynamical systems \cite{Seydel:1994}. There are also equivalents to Lyapunov exponents in RBNs \cite{LuqueSole2000}. The rationale is similar in all of these methods: if perturbations do not propagate, then the system is in the ordered dynamical phase. If perturbations propagate through the system, then it is in the chaotic dynamical phase. The phase transition (critical regime) lies  where the size of the perturbation remains constant in time. 

To measure statistically the sensitivity to initial conditions of MRBNs, we used the following method \cite{Gershenson2004b}: For a randomly generated network, pick a random initial state $S_i$, and let it run for a large number of steps $t_{max}$ ($t_{max}=10000$ in our present experiments), to reach a final state $S_f$. Now, apply a random point ``mutation" to initial state $S_i$ to obtain $S'_i$, i.e. do a random bit flip. Then, let the network run for $t_{max}$ from $S'_i$ to obtain another final state $S'_f$. The difference between states can be calculated with the normalized Hamming distance:

\begin{equation}
H(A,B)=\frac{1}{N_T}\sum\limits_{i}^{N_T}\left\vert a_{i}-b_{i}\right\vert
\end{equation}

If states $A$ and $B$ are equal, then $H(A,B)=0$. The maximum $H=1$ is given when $A$ is the complementary state of $B$, i.e. every node with state one in $A$ has a state zero in $B$ and every node with state zero in $A$ has a state one in $B$, i.e. full anticorrelation. $H=0.5$ implies no correlation between $A$ and $B$. The smaller $H$ is, the more similar $A$ and $B$ are. As $H$ increases (up to $H=0.5$), it implies that differences between $A$ and $B$ also increase.

Since there is only one bit difference between $S_i$ and $S'_i$ and each state has $N_T$ bits:

\begin{equation}
H_i=H(S_i,S'_i)=\dfrac{1}{N_T}
\end{equation}

Now, to measure the sensitivity to initial conditions, the difference between the final and initial Hamming distances $\Delta H$ is used:

\begin{equation}
\Delta H=H_f-H_i
\end{equation}
where
\begin{equation}
H_f=H(S_f,S'_f)
\end{equation}

A large number of random initial states for a large number of MRBNs are used to calculate an average $\Delta H$ for an ensemble. 

If $\Delta H<0$, then different initial states converge to the same final state. This is a characteristic of the ordered regime, where trajectories in state space tend to converge. If $\Delta H>0$, then small differences in initial states tend to increase, a characteristic of the chaotic regime, where trajectories in state space tend to diverge. If $\Delta H=0$, them small initial differences are maintained, a characteristic of the critical regime, where trajectories in state space neither converge nor diverge (in practice, $\Delta H\approx 0$).
Thus, the average $\Delta H$ can indicate the regime of a MRBN.

\begin{figure}[ht!]
\begin{center}
\includegraphics[width=0.8\textwidth]{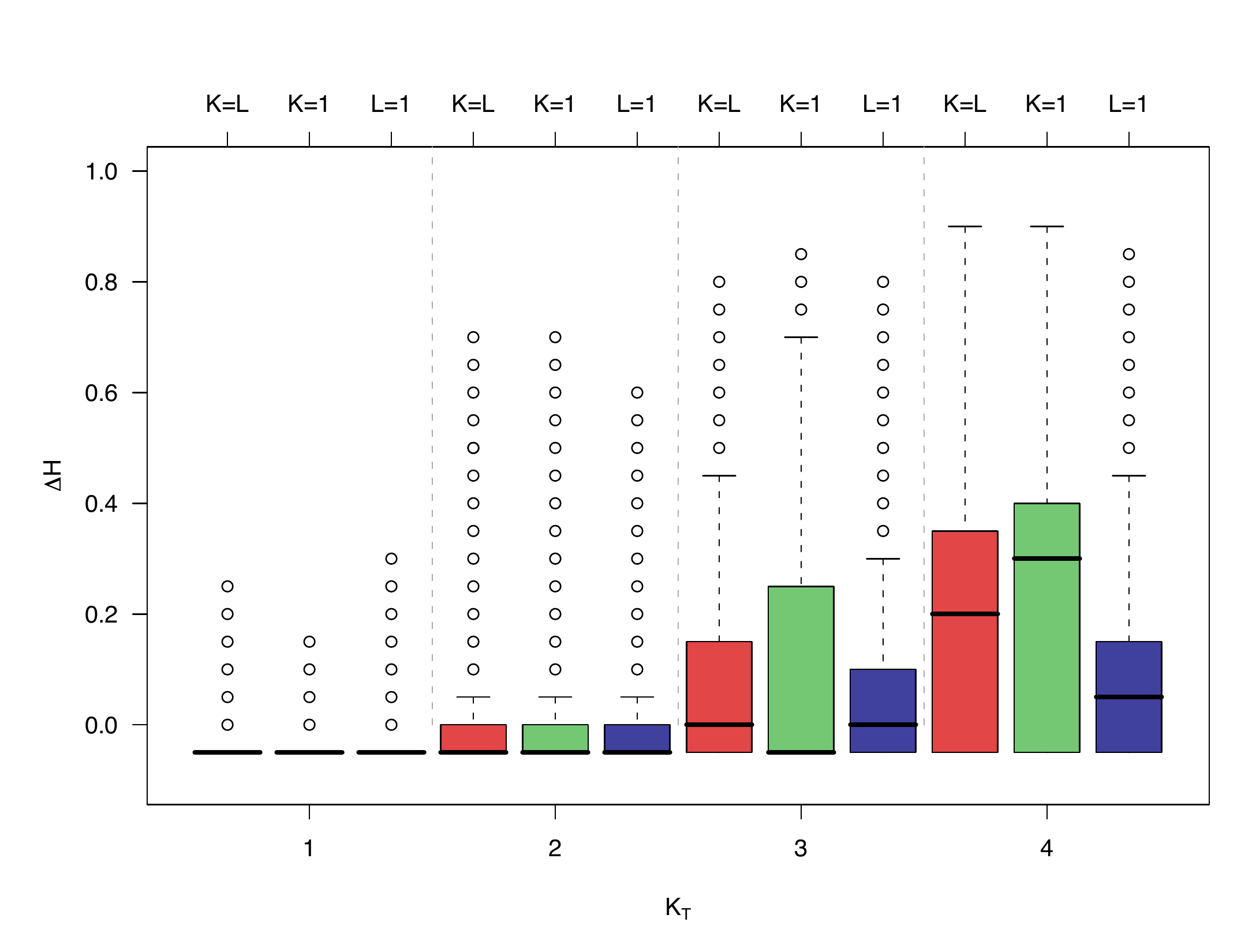}
\end{center}
 \caption{Sensitivity to initial conditions for cases 1, 2 \& 3 ($K=L$, $K=1$, and $L=1$, respectively).}
\label{fig:phase1}
\end{figure}

Boxplots of the results for cases 1, 2 and 3 are shown in Figure \ref{fig:phase1}. Notice that boxplots show medians. Means can be compared in Tables \ref{tab:phasec1}, \ref{tab:phasec2}, and \ref{tab:phasec3}; found in Appendix \ref{sec:tables}.

For $K_T=1$, the dynamics are in the ordered regime for all cases, since small differences in initial states tend to be reduced, indicated by a negative $\Delta H = -\frac{1}{N_T}$. 
For $K_T=2$, Tables \ref{tab:phasec1}, \ref{tab:phasec2}, and \ref{tab:phasec3} in Appendix \ref{sec:tables} show that the average $\Delta H$ is close to zero for all cases, suggesting a critical regime. The difference caused by modularity is clearly seen for $K_T>2$, i.e. in the chaotic regime. The sensitivity to initial conditions is inversely correlated with $\kappa$. This can be explained as follows: the higher the $\kappa$, the more ``isolated" modules are. Thus, it is more difficult for damage to spread between modules, even when average connectivities $K_T$  are high. It can be said that modularity in MRBNs brings the dynamics of the chaotic regime closer to criticality. Notice that the phase transition ($K_T=2$, $\Delta H=0$) does not move with $\kappa$. Rather, the properties of the critical regime are expanded into the chaotic regime by higher values of $\kappa$.

\begin{figure}[ht!]
\begin{center}
\includegraphics[width=0.8\textwidth]{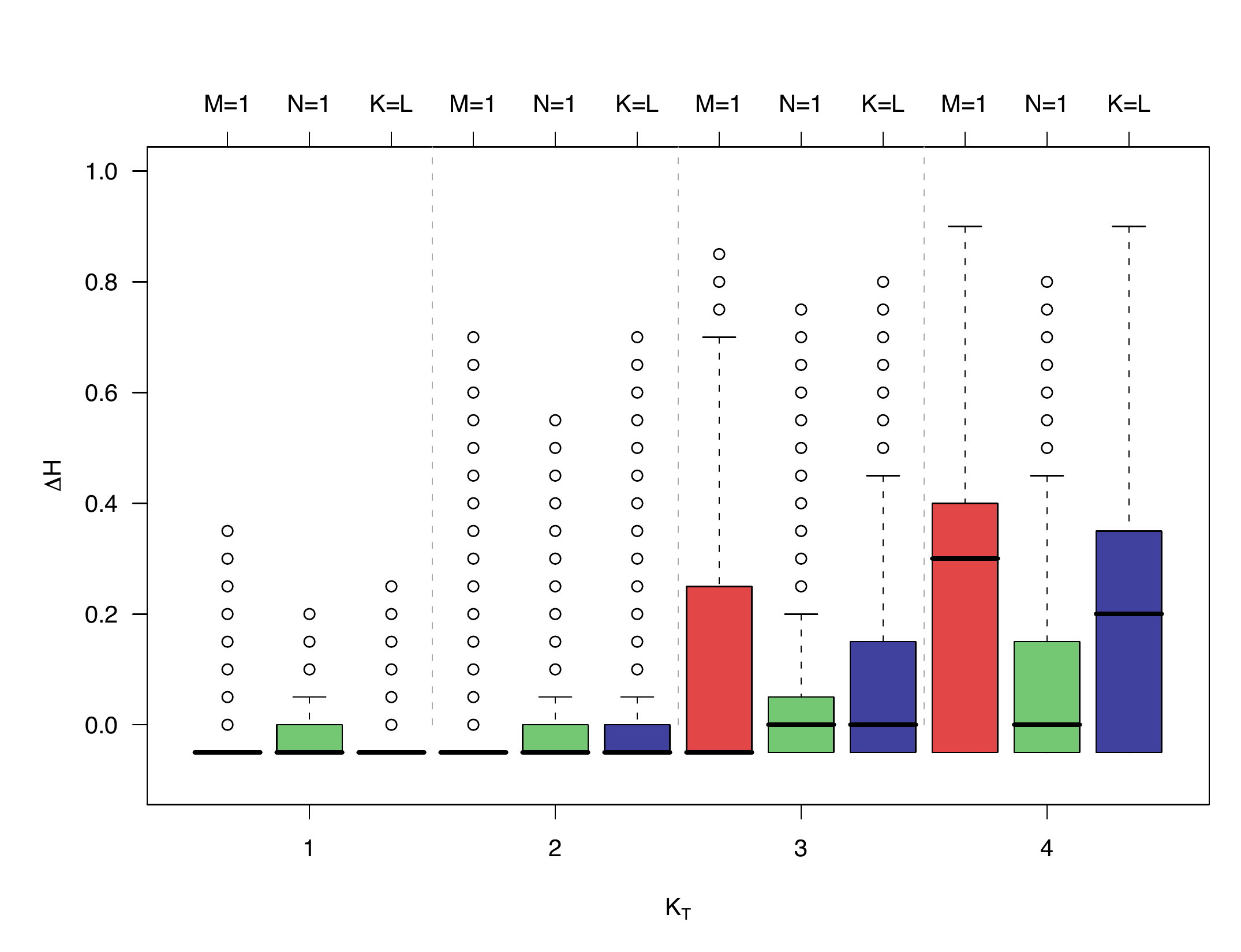}
\end{center}
 \caption{Sensitivity to initial conditions for cases 4, 5 \& 1 ($M=1$, $N=1$, and $K=L$, respectively).}
\label{fig:phase2}
\end{figure}

From Figure \ref{fig:phase2} it can be seen that $\mu$ is also relevant for the sensitivity to initial conditions within the chaotic regime ($K_T>2$). Case 4 (equivalent to a classical RBN) has the highest sensitivity, since damage can equally spread among all nodes in the MRBN. The extreme case 5 has a very low sensitivity to initial conditions. On the one hand, it is because of the fictitious links already explained. On the other hand, since all nodes have self-connections when $K\geq 1$, damage will have a lower probability of spreading to other nodes (modules). The intermediate case 1 shows that some modularity prevents damage from spreading between modules, bringing the chaotic dynamics closer to the critical regime. 

\subsubsection{Larger Networks}

\begin{figure}[ht!]
\begin{center}
\includegraphics[width=0.6\textwidth]{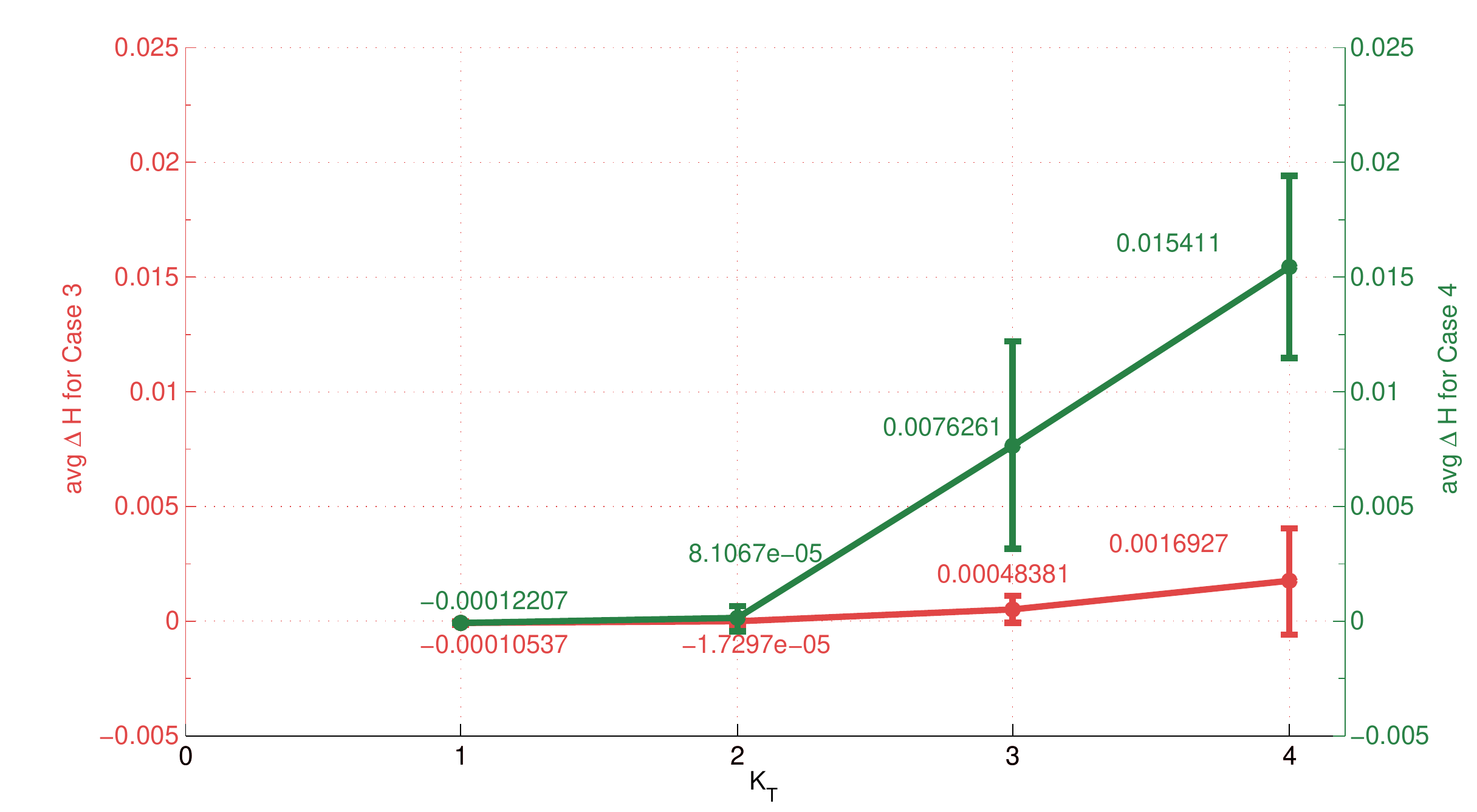}
\end{center}
 \caption{Mean sensitivity to initial conditions for cases 3 \& 4 ($M=N, L=1$ and $M=1, L=0$, respectively) for large networks ($N_T=400$). Error bars indicate standard deviations.}
\label{fig:phase400}
\end{figure}

To ensure that the results presented above were not an artifact of the small size of the networks ($N_T=20$)  specific experiments were performed to compare cases 3 and 4 with large networks of $N_T=400$. 
For case 3, $N=M=20$ and $L=1$. For case 4, $M=1$, $N=N_T=400$, $K=K_T$, and $L=0$.
For each MRBN family, only one hundred networks were generated and only one hundred state pairs were explored for ten thousand steps. These experiments are less statistically significant, but they clearly show that the difference in sensitivity to initial conditions is due to modularity and not to network size.
Results can be observed in Figure \ref{fig:phase400}. It can be seen that the advantage of modularity is increased for larger networks.

\section{Discussion}
\label{sec:discussion}

In the previous section, the statistical results showed that MRBNs are more robust than classic RBNs for the same $K_T$ values, since modularity reduces the probability of damage to spread. This can also be confirmed analytically.

The method of Luque and Sol\'{e} \cite{LuqueSole1997} can be extended to MRBNs. Instead of focussing on a single node with a Boolean state, one can focus on a single module $m$ with internal dynamics. The probability that the internal dynamics of $m$ may be perturbed will depend mainly on $\avg{L}$ (as well as $p$). If $p=0.5$, then $L<2$ implies that on average damage will not propagate between modules, even if the internal dynamics of $m$ are chaotic, i.e. $K>2$. Still, if $K>>2$, i.e. $m$ is deep within the chaotic phase because of a high number of intramodular links, then the fragility of the network will be noticeable in the MRBN ($\Delta H >>0$) even if damage does not spread outside $m$. It is clear that damage spreading will depend on the value of $K$ as well, e.g. if $K$ is small, damage will have a lower probability of propagating within modules, affecting the probability of spread across modules.

Since damage across the whole MRBN can spread through internal ($K$) or external ($L$) links, it can be seen that there is the following relationship between critical $L$ and $K$, extending equation \ref{eq:kc}:

\begin{equation}
\avg{L_c}\avg{K_c} = \frac{1}{2p(1-p)}
\label{eq:lkc}
\end{equation}
\begin{equation}
\avg{L_c} = \frac{1}{2\avg{K_c}p(1-p)}
\label{eq:lcp}
\end{equation}
and for $p=0.5$:

\begin{equation}
\avg{L_c} = \frac{2}{\avg{K_c}}
\label{eq:lc}
\end{equation}
and
\begin{equation}
\avg{K_c} = \frac{2}{\avg{L_c}}
\label{eq:lc}
\end{equation}

\begin{figure}[ht!]
\begin{center}
\includegraphics[width=0.75\textwidth]{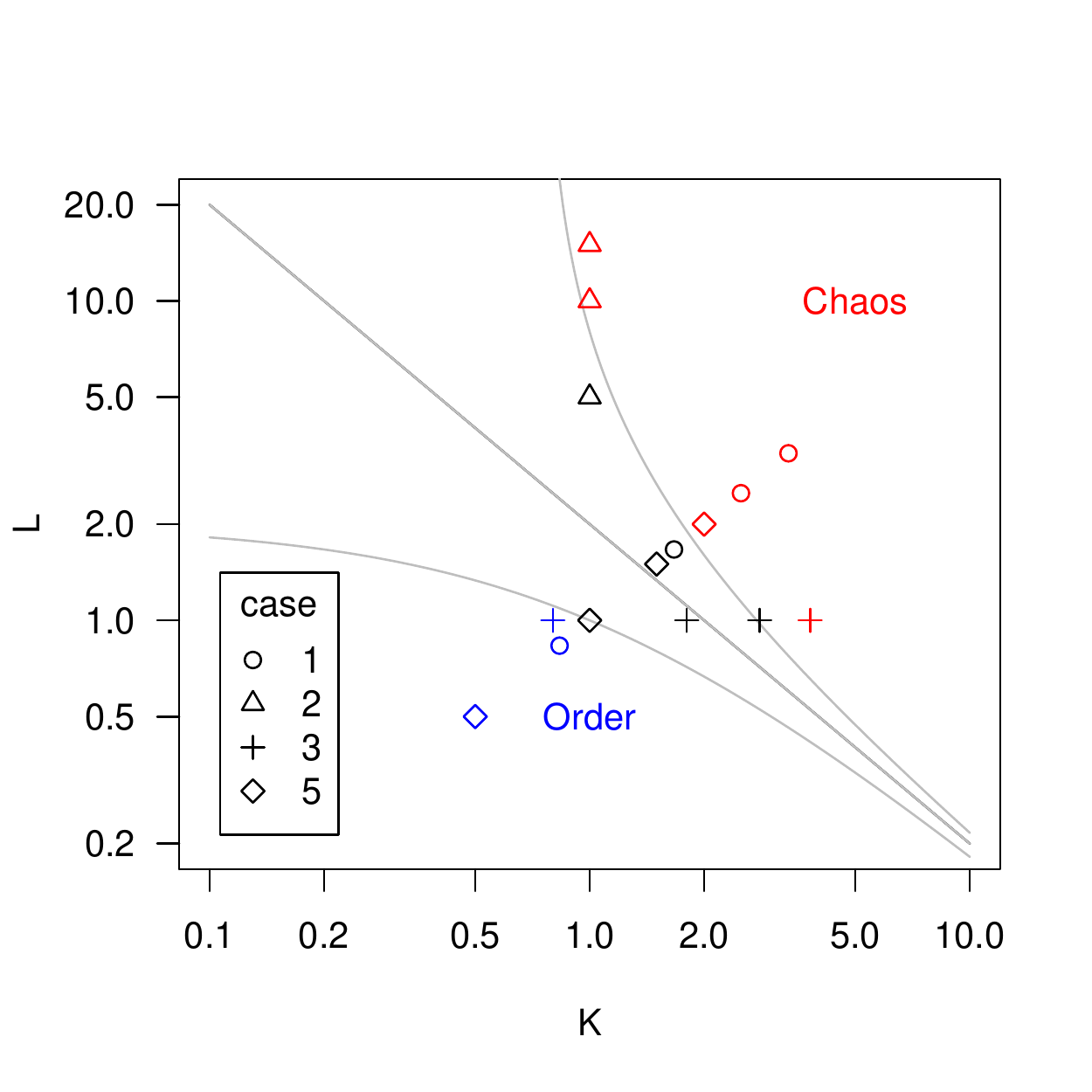}
\end{center}
 \caption{Theoretical criticality of MRBNs (line) depending on $K$ and $L$. Above the line there is the chaotic phase, while the ordered phase lies below the line. Notice logarithmic scales. Experimental averages are shown, except when $L=0$ (See Tables \ref{tab:phasec1}--\ref{tab:phasec5}). Different shapes denote different cases (see legend), while different colors denote different regimes, obtained from $\Delta H$ (blue for ordered, black for critical, red for chaotic; color in online version). Due to finite-size effects ($N_T=20$), criticality is considered when $-0.0333 < \Delta H < 0.0333$. The experimental results that are considered as critical can be fitted between the curves $L=\frac{2}{K-0.75}$ and $L=\frac{2}{K+1}$, shown in gray. The deviations from $L=\frac{2}{K}$ are also due to finite sized effects. Results for larger networks are closer to $L=\frac{2}{K}$ and to $\Delta H=0$.}
\label{fig:lc}
\end{figure}

A plot of equation \ref{eq:lc} can be seen in Figure \ref{fig:lc}, with simulation averages for cases 1, 2, 3, and 5\footnote{Case 4 is omitted because there is only one module, so the results are not related to how damage propagates across modules. Moreover, $L=0$ for case 4.}. Even when the analysis presented above assumes infinitely-sized modules and networks, the simulation results with small networks match the theoretical analysis.

It can be seen that a higher $\kappa$ implies lower $L$, i.e. lower values in Figure \ref{fig:lc}. When  $\kappa=1$, $L=0$, damage cannot propagate between modules, even for the highest local connectivities ($K=N$). Thus, in principle $\kappa$ can be used to modulate the sensitivity to initial conditions of MRBNs \cite{Gershenson:2010}.

There is a negative correlation between $\kappa$ and $\Delta H$, and a positive correlation between $\kappa$ and $A$. However, the explanations for the effect on sensitivity to initial conditions and on the number of attractors seem to be also related to different effects of the topology of MRBNs. Still, it would be interesting to study whether it is always the case that RBNs with more attractors on average are more robust to damage spread, as it is the case for MRBNs. Alternatively, finding counterexamples would be illustrative as well.

\section{Conclusions and Future Work}

We have presented a generalization of random Boolean networks, where modules can be constructed. With statistical studies on ensembles of MRBNs and an analytical study, it could be seen that different parameters that define MRBNs affect most of the properties of the networks and their dynamics. Thus, it can be said that modularity is one way of guiding the self-organization of RBNs \cite{Gershenson:2010}.

A drawback of the studies of DDN criticality based on sensitivity to initial conditions is that it restricts the critical regime to a phase transition. 
Information-theoretical measures \cite{Lizier:2008,Wang:2010} might offer an alternative to better characterize the critical regime and the effect of modularity and other properties on criticality.

It was shown here that a high $\kappa$ (high percentage of internal inputs) and a low $\mu$ (node-to-module ratio) promote criticality in what otherwise would be a chaotic regime. 
However, is there an ``optimal" value of $\kappa$ and/or $\mu$ for particular systems? It would be also interesting to measure the modularity of real GRNs, and measure to what extent the modularity plays a role in their criticality \cite{Balleza:2008}. Modularity might help explain why real GRNs tend to have a high average connectivity that would set them in the chaotic regime \cite{Harris:2002}, while exhibiting critical behavior \cite{Balleza:2008}.

The results presented here are encouraging to study modularity at multiple scales, i.e. nested modules or hierarchies. RBNs have two scales: nodes and network, with no modularity. MRBNs have three scales: nodes, modules, and networks. The MRBN model can be generalized to have an arbitrary number of intermediate scales, i.e. modules of modules, with different coupling preferences. In this way, a multi-scaled modular network could have in principle different dynamical regimes at different scales, i.e. damage could propagate at one scale but not at another. A general way of defining multiple scales might be with recursive RBNs.

A scale-free topology has been shown to promote criticality in what otherwise would be ordered networks \cite{Aldana2003}. Scale-free topology and criticality are also present in natural networks \cite{AldanaCluzel2003,Oikonomou:2006}.
It would be interesting to study how modular and scale free topologies could be combined, and whether their effects on criticality are cummulative: for abstract RBNs and for real GRNs. For example, in our present studies $N$ is constant for all modules, while the module size could have a scale-free distribution (few large modules and several small modules).

The redundancy of nodes \cite{GershensonEtAl2006} and modules \cite{Benitez:2010} has shown to promote robustness in RBNs. MRBNs can be general models to study the relationship between modularity, robustness, evolvability, and criticality. Also, degeneracy can play an important role in robustness \cite{Wagner2004,Wagner2005} and evolvability \cite{Whitacre:2010}, although the role of degeneracy in the criticality of RBNs still remains to be explored. 

The potential avenues of research are several. The topics related to modularity and criticality are many. We believe that MRBNs can contribute to illuminate these interesting questions.

\section*{Acknowledgments}

We should like to thank Joseph Lizier, Rosalind Wang, and Borys Wrobel for useful comments and suggestions. 
R.P.B. was supported by CONACyT scholarship 268628 and PAEP, UNAM. 
C.G. was partially supported by SNI membership 47907 of CONACyT, Mexico.

\bibliographystyle{alj}
\bibliography{carlos,rbn,sos,complex,evolution}

\begin{thebibliography}{10}

\bibitem{Aldana2003}
Aldana, M. (2003) Boolean dynamics of networks with scale-free topology. {\em
  Physica D\/}, {\bf 185}, 45--66.

\bibitem{AldanaCluzel2003}
Aldana, M. and Cluzel, P. (2003) {A natural class of robust networks}. {\em
  Proceedings of the National Academy of Sciences of the United States of
  America\/}, {\bf 100}, 8710--8714.

\bibitem{AldanaEtAl2003}
Aldana-Gonz{\'a}lez, M., Coppersmith, S., and Kadanoff, L.~P. (2003) Boolean
  dynamics with random couplings. Kaplan, E., Marsden, J.~E., and Sreenivasan,
  K.~R. (eds.), {\em Perspectives and Problems in Nonlinear Science. A
  Celebratory Volume in Honor of Lawrence Sirovich\/}, Springer Applied
  Mathematical Sciences Series.

\bibitem{andrecut05}
Andrecut, M. (2005) Mean field dynamics of random {Boolean} networks. {\em
  Journal of Statistical Mechanics\/}, {\bf P02003}.

\bibitem{Balleza:2008}
Balleza, E., Alvarez-Buylla, E.~R., Chaos, A., Kauffman, S., Shmulevich, I.,
  and Aldana, M. (2008) Critical dynamics in genetic regulatory networks:
  Examples from four kingdoms. {\em PLoS ONE\/}, {\bf 3}, e2456.

\bibitem{Bastolla:1998}
Bastolla, U. and Parisi, G. (1998) {The modular structure of Kauffman
  networks}. {\em Physica D: Nonlinear Phenomena\/}, {\bf 115}, 219--233.

\bibitem{Benitez:2010}
Ben{\'\i}tez, M. and Alvarez-Buylla, E.~R. (2010) Dynamic-module redundancy
  confers robustness to the gene regulatory network involved in hair patterning
  of \emph{Arabidopsis} epidermis. {\em Biosystems\/}, {\bf 102}, 11--15.

\bibitem{callebaut05}
{Callebaut}, W. and {Rasskin-Gutman}, D. (2005) {\em Modularity: Understanding
  the Development and Evolution of Natural Complex Systems\/}. Vienna Series in
  Theoretical Biology, {The MIT Press}.

\bibitem{Crutchfield:1994}
Crutchfield, J. (1994) {Critical computation, phase transitions, and
  hierarchical learning}. Yamaguti, M. (ed.), {\em Towards the Harnessing of
  Chaos\/}, pp. 29--46, Elsevier.

\bibitem{Csermely:2006}
Csermely, P. (2006) {\em Weak Links: Stabilizers of Complex Systems from
  Proteins to Social Networks\/}. Springer.

\bibitem{Damiani:2010}
Damiani, C., Kauffman, S., Serra, R., Villani, M., and Colacci, A. (2010)
  Information transfer among coupled random {Boolean} networks. Bandini, S.,
  Manzoni, S., Umeo, H., and Vizzari, G. (eds.), {\em Cellular Automata\/},
  vol. 6350 of {\em Lecture Notes in Computer Science\/}, pp. 1--11, Springer
  Berlin / Heidelberg.

\bibitem{DerridaPomeau1986}
Derrida, B. and Pomeau, Y. (1986) Random networks of automata: A simple
  annealed approximation. {\em Europhys. Lett.\/}, {\bf 1}, 45--49.

\bibitem{espinoza-soto}
{Espinoza-Soto}, C. and {Wagner}, A. (2010) Specialization can drive the
  evolution of modularity. {\em PLoS Computational Biology\/}, {\bf 6},
  e1000719.

\bibitem{FernandezSole2003}
Fern{\'a}ndez, P. and Sol{\'e}, R. (2004) The role of computation in complex
  regulatory networks. Koonin, E.~V., Wolf, Y.~I., and Karev, G.~P. (eds.),
  {\em Power Laws, Scale-Free Networks and Genome Biology\/}, Landes
  Bioscience.

\bibitem{Gershenson2002e}
Gershenson, C. (2002) Classification of random {Boolean} networks. Standish,
  R.~K., Bedau, M.~A., and Abbass, H.~A. (eds.), {\em Artificial Life {VIII}:
  Proceedings of the Eight International Conference on Artificial Life\/}, pp.
  1--8, MIT Press.

\bibitem{Gershenson2004c}
Gershenson, C. (2004) Introduction to random {Boolean} networks. Bedau, M.,
  Husbands, P., Hutton, T., Kumar, S., and Suzuki, H. (eds.), {\em Workshop and
  Tutorial Proceedings, Ninth International Conference on the Simulation and
  Synthesis of Living Systems {(ALife} {IX)}\/}, Boston, MA, pp. 160--173.

\bibitem{Gershenson2004b}
Gershenson, C. (2004) Updating schemes in random {Boolean} networks: Do they
  really matter? Pollack, J., Bedau, M., Husbands, P., Ikegami, T., and Watson,
  R.~A. (eds.), {\em Artificial Life {IX} Proceedings of the Ninth
  International Conference on the Simulation and Synthesis of Living
  Systems\/}, pp. 238--243, MIT Press.

\bibitem{RBNLab}
Gershenson, C. (2005), {RBNLab}. {http://rbn.sourceforge.net}.

\bibitem{Gershenson:2010}
Gershenson, C. (In Press) Guiding the self-organization of random {Boolean}
  networks. {\em Theory in Biosciences\/}.

\bibitem{GershensonEtAl2006}
Gershenson, C., Kauffman, S.~A., and Shmulevich, I. (2006) The role of
  redundancy in the robustness of random {Boolean} networks. Rocha, L.~M.,
  Yaeger, L.~S., Bedau, M.~A., Floreano, D., Goldstone, R.~L., and Vespignani,
  A. (eds.), {\em Artificial Life X, Proceedings of the Tenth International
  Conference on the Simulation and Synthesis of Living Systems.\/}, pp. 35--42,
  MIT Press.

\bibitem{Harris:2002}
Harris, S.~E., Sawhill, B.~K., Wuensche, A., and Kauffman, S. (2002) A model of
  transcriptional regulatory networks based on biases in the observed
  regulation rules. {\em Complexity\/}, {\bf 7}, 23--40.

\bibitem{ho05}
{Ho}, M., {Hung}, Y., and {Jiang}, I. (2005) Stochastic coupling of two random
  {Boolean} networks. {\em Physics Letters A\/}, {\bf 344}, 36--42.

\bibitem{HuangIngber2000}
Huang, S. and Ingber, D.~E. (2000) Shape-dependent control of cell growth,
  differentiation, and apoptosis: Switching between attractors in cell
  regulatory networks. {\em Exp. Cell Res.\/}, {\bf 261}, 91--103.

\bibitem{hung06}
Hung, Y.-C., Ho, M.-C., Lih, J.-S., and Jiang, I.-M. (2006) Chaos
  synchronization of two stochastically coupled random {Boolean} networks. {\em
  Physics Letters A\/}, {\bf 356}, 35 -- 43.

\bibitem{Kauffman1969}
Kauffman, S.~A. (1969) Metabolic stability and epigenesis in randomly
  constructed genetic nets. {\em Journal of Theoretical Biology\/}, {\bf 22},
  437--467.

\bibitem{Kauffman1993}
Kauffman, S.~A. (1993) {\em The Origins of Order\/}. Oxford University Press.

\bibitem{Kauffman2000}
Kauffman, S.~A. (2000) {\em Investigations\/}. Oxford University Press.

\bibitem{Kauffman2004}
Kauffman, S.~A. (2004) The ensemble approach to understand genetic regulatory
  networks. {\em Physica A: Statistical Mechanics and its Applications\/}, {\bf
  340}, 733--740.

\bibitem{Kauffman:2004}
Kauffman, S.~A., Peterson, C., Samuelsson, B., and Troein, C. (2004) {Genetic
  networks with canalyzing Boolean rules are always stable}. {\em Proceedings
  of the National Academy of Sciences of the United States of America\/}, {\bf
  101}, 17102--17107.

\bibitem{Langton1990}
Langton, C. (1990) Computation at the edge of chaos: Phase transitions and
  emergent computation. {\em Physica D\/}, {\bf 42}, 12--37.

\bibitem{Lizier:2008}
Lizier, J., Prokopenko, M., and Zomaya, A. (2008) The information dynamics of
  phase transitions in random {Boolean} networks. Bullock, S., Noble, J.,
  Watson, R., and Bedau, M.~A. (eds.), {\em Artificial Life XI - Proceedings of
  the Eleventh International Conference on the Simulation and Synthesis of
  Living Systems\/}, pp. 374--381, MIT Press.

\bibitem{LuqueSole1997}
Luque, B. and Sol{\'e}, R.~V. (1997) Phase transitions in random networks:
  Simple analytic determination of critical points. {\em Physical Review E\/},
  {\bf 55}, 257--260.

\bibitem{LuqueSole2000}
Luque, B. and Sol{\'e}, R.~V. (2000) Lyapunov exponents in random {Boolean}
  networks. {\em Physica A\/}, {\bf 284}, 33--45.

\bibitem{Oikonomou:2006}
Oikonomou, P. and Cluzel, P. (2006) Effects of topology on network evolution.
  {\em Nature Physics\/}, {\bf 2}, 532--536.

\bibitem{Balpo:2011}
{Poblanno-Balp}, R. (2011) {\em Coupled Random {Boolean} Networks and Their
  Criticality\/}. Master's thesis, Universidad Nacional Aut\'onoma de M\'exico,
  Ciudad Universitaria, M\'exico.

\bibitem{BalpoGershenson:2010}
{Poblanno-Balp}, R. and Gershenson, C. (2010) Modular random {Boolean}
  networks. Fellermann, H., D\"{o}rr, M., Hanczyc, M.~M., Laursen, L.~L.,
  Maurer, S., Merkle, D., Monnard, P.-A., St$\o$y, K., and Rasmussen, S.
  (eds.), {\em {Artificial Life XII} Proceedings of the Twelfth International
  Conference on the Synthesis and Simulation of Living Systems\/}, pp.
  303--304, MIT Press.

\bibitem{schlosser}
{Schlosser}, G. and {Wagner}, G.~P. (2004) {\em Modularity in Development and
  Evolution\/}. The University of Chicago Press.

\bibitem{segal}
{Segal}, E., {Shapira}, M., {Regev}, A., {Pe'er}, D., {Botstein}, D., {Koller},
  D., and {Friedman}, N. (2003) Module networks: identifying regulatory modules
  and their condition-specific regulators from gene expression data. {\em
  Nature Genetics\/}, {\bf 34}, 166--176.

\bibitem{serra08}
Serra, R., Villani, M., Damiani, C., Graudenzi, A., and Colacci, A. (2008) The
  diffusion of perturbations in a model of coupled random boolean networks.
  Umeo, H., Morishita, S., Nishinari, K., Komatsuzaki, T., and Bandini, S.
  (eds.), {\em Cellular Automata\/}, vol. 5191 of {\em Lecture Notes in
  Computer Science\/}, pp. 315--322, Springer Berlin -- Heidelberg.

\bibitem{Seydel:1994}
Seydel, R. (1994) {\em {Practical bifurcation and stability analysis: from
  equilibrium to chaos}\/}. Springer.

\bibitem{Shmulevich:2004}
Shmulevich, I. and Kauffman, S.~A. (2004) Activities and sensitivities in
  {Boolean} network models. {\em Phys. Rev. Lett.\/}, {\bf 93}, 048701.

\bibitem{Simon1996}
Simon, H.~A. (1996) {\em The Sciences of the Artificial\/}. MIT Press, 3rd edn.

\bibitem{Stauffer:1994}
Stauffer, D. (1994) Evolution by damage spreading in {Kauffman} model. {\em
  Journal of Statistical Physics\/}, {\bf 74}, 1293--1299.

\bibitem{Thomas1978}
Thomas, R. (1978) Logical analysis of systems comprising feedback loops. {\em
  J. Theor. Biol.\/}, {\bf 73}, 631--656.

\bibitem{villani06}
{Villani}, M., {Serra}, R., {Kauffman}, S.~A., and {Ingrami}., P. (2006)
  Coupled random {Boolean} network forming an artificial tissue. {\em Cellular
  Automata\/}, vol. 4173 of {\em Lecture Notes in Computer Science\/}, pp.
  548--556, Springer.

\bibitem{Wagner2004}
Wagner, A. (2005) Distributed robustness versus redundancy as causes of
  mutational robustness. {\em BioEssays\/}, {\bf 27}, 176--188.

\bibitem{Wagner2005}
Wagner, A. (2005) {\em Robustness and Evolvability in Living Systems\/}.
  Princeton University Press.

\bibitem{wagner07}
Wagner, G.~P., Pavlicev, M., and Cheverud, J.~M. (2007) The road to modularity.
  {\em Nature reviews. Genetics\/}, {\bf 8}, 921--931.

\bibitem{Wang:2010}
Wang, X.~R., Lizier, J., and Prokopenko, M. (2010) A {Fisher} information study
  of phase transitions in random {Boolean} networks. Fellermann, H., D\"{o}rr,
  M., Hanczyc, M.~M., Laursen, L.~L., Maurer, S., Merkle, D., Monnard, P.-A.,
  St$\o$y, K., and Rasmussen, S. (eds.), {\em {Artificial Life XII} Proceedings
  of the Twelfth International Conference on the Synthesis and Simulation of
  Living Systems\/}, pp. 305--312, MIT Press.

\bibitem{Watson:2006}
Watson, R.~A. (2006) {\em Compositional Evolution: The Impact of Sex,
  Symbiosis, and Modularity on the Gradualist Framework of Evolution\/}. MIT
  Press.

\bibitem{Whitacre:2010}
Whitacre, J.~M. and Bender, A. (2010) Degeneracy: a design principle for
  robustness and evolvability. {\em Journal of Theoretical Biology\/}, {\bf
  263}, 143--153.

\bibitem{Wolfram1986}
Wolfram, S. (1986) {\em Theory and Application of Cellular Automata\/}. World
  Scientific.

\bibitem{Wuensche1998}
Wuensche, A. (1998) Discrete dynamical networks and their attractor basins.
  Standish, R., Henry, B., Watt, S., Marks, R., Stocker, R., Green, D., Keen,
  S., and Bossomaier, T. (eds.), {\em Complex Systems '98\/}, University of New
  South Wales, Sydney, Australia, pp. 3--21.

\end{thebibliography}

\appendix

\section{Tables}

\renewcommand{\thetable}{\thesection.\arabic{table}}
\setcounter{table}{0}

\label{sec:tables}


\begin{table}[ht!]
\setlength{\tabcolsep}{1.8pt}
	\begin{center}
	\caption[]{Statistical results for case 1: $K = L$, \quad $\mu\rightarrow1$.}
\label{tab:statsc1}
		{ \small \begin{tabular}{ c | c | c | c | c | c | c | c | c | c | c | c | c}
			\hline \hline
			\textbf{N} & \textbf{K=L} & \textbf{M} & \textbf{T} & \textbf{$K_T$} & \textbf{A} & \textbf{Le} & \textbf{\%SIA} & $\mu$ & $\kappa$ & $\sigma A$ & $\sigma Le$ & $\sigma \%SIA$ \\
			\hline \hline
			5 & 0.83333 & 4 & 20 & 1 & 4.03 & 2 & 0.0009 & 1.25 & 0.83333 & 5.722 & 1.435 & 0.0019 \\ \hline
			5 & 1.66666 & 4 & 40 & 2 & 12.65 & 4 & 0.0051 & 1.25 & 0.83333 & 16.073 & 3.619 & 0.0083 \\ \hline
			5 & 2.5 & 4 & 60 & 3 & 18.06 & 7 & 0.0106 & 1.25 & 0.83333 & 18.618 & 6.140 & 0.0122 \\ \hline
			5 & 3.33333 & 4 & 80 & 4 & 14.10 & 13 & 0.0125 & 1.25 & 0.83333 & 14.789 & 13.468 & 0.0129 \\ \hline
		\end{tabular}}
	\end{center}
\end{table}

\begin{table}[!ht]
	\setlength{\tabcolsep}{1.8pt}
	\begin{center}
	\caption[]{Statistical results for case 2: $K = 1$,\quad$\mu\rightarrow1$.}
\label{tab:statsc2}
		{ \small \begin{tabular}{ c | c | c | c | c | c | c | c | c | c | c | c | c | c}
			\hline \hline
			\textbf{N} & \textbf{K} & \textbf{M} & \textbf{L} & \textbf{T} & \textbf{$K_T$} & \textbf{A} & \textbf{Le} & \textbf{\%SIA} & $\mu$ & $\kappa$ & $\sigma A$ & $\sigma L$ & $\sigma \%SIA$ \\
			\hline \hline
			5 & 1 & 4 & 0 & 20 & 1 & 5.38 & 2 & 0.0013 & 1.25 & 1 & 8.063 & 1.848 & 0.0025 \\ \hline
			5 & 1 & 4 & 5 & 40 & 2 & 4.89 & 3 & 0.0016& 1.25 & 0.5 & 5.045 & 3.012 & 0.0026 \\ \hline
			5 & 1 & 4 & 10 & 60 & 3 & 4.66 & 7 & 0.0028& 1.25 & 0.33333 & 3.583 & 7.573 & 0.0029 \\ \hline
			5 & 1 & 4 & 15 & 80 & 4 & 4.39 & 22 & 0.0079& 1.25 & 0.25 & 2.312 & 26.448 & 0.0080 \\ \hline
		\end{tabular}}
	\end{center}
\end{table}

\begin{table}[!ht]
	\setlength{\tabcolsep}{1.8pt}
	\begin{center}
	\caption[]{Statistical results for case 3: $L = 1$,\quad$\mu\rightarrow1$.}
\label{tab:statsc3}
		{ \small \begin{tabular}{ c | c | c | c | c | c | c | c | c | c | c | c | c | c}
			\hline \hline
			\textbf{N} & \textbf{K} & \textbf{M} & \textbf{L} & \textbf{T} & \textbf{$K_T$} & \textbf{A} & \textbf{Le} & \textbf{\%SIA} & $\mu$ & $\kappa$ & $\sigma A$ & $\sigma L$ & $\sigma \%SIA$ \\
			\hline \hline
			5 & 0.8 & 4 & 1 & 20 & 1 & 3.45 & 2 & 0.0007 & 1.25	& 0.8 & 4.155 & 1.508 & 0.0013 \\ \hline
			5 & 1.8 & 4 & 1 & 40 & 2 & 15.77 & 5 & 0.0071 & 1.25	& 0.9 & 20.528 & 3.792 & 0.0123 \\ \hline
			5 & 2.8 & 4 & 1 & 60 & 3 & 29.30 & 10 & 0.0231 & 1.25	& 0.933333 & 28.898 & 10.294 & 0.0338 \\ \hline
			5 & 3.8 & 4 & 1 & 80 & 4 & 34.36 & 14 & 0.0376 & 1.25	& 0.95 & 31.973 & 14.615 & 0.0432 \\ \hline
		\end{tabular}}
	\end{center}
\end{table}

\begin{table}[!ht]
	\setlength{\tabcolsep}{1.8pt}
	\begin{center}
	\caption[]{Statistical results for case 4: $M = 1$, \quad $N = 20$,\quad$L = 0$\quad, \quad $K=K_T$.}
\label{tab:statsc4}
		{ \small \begin{tabular}{ c | c | c | c | c | c | c | c | c | c | c | c | c | c}
			\hline \hline
			\textbf{N} & \textbf{K} & \textbf{M} & \textbf{L} & \textbf{T} & \textbf{$K_T$} & \textbf{A} & \textbf{Le} & \textbf{\%SIA} & $\mu$ & $\kappa$ & $\sigma A$ & $\sigma L$ & $\sigma \%SIA$ \\
			\hline \hline
			20 & 1 & 1 & 0 & 20 & 1 & 1.68 & 2 & 0.0003 & 20 & 1 & 1.667 & 1.997 & 0.0011 \\ \hline
			20 & 2 & 1 & 0 & 40 & 2 & 3.15 & 3 & 0.0010 & 20 & 1 & 3.371 & 2.762 & 0.0016 \\ \hline
			20 & 3 & 1 & 0 & 60 & 3 & 4.23 & 8 & 0.0026 & 20 & 1 & 3.563 & 9.586 & 0.0027 \\ \hline
			20 & 4 & 1 & 0 & 80 & 4 & 4.43 & 22 & 0.0081 & 20 & 1 & 2.407 & 23.662 & 0.0078 \\ \hline
		\end{tabular}}
	\end{center}
\end{table}

\begin{table}[!ht]
	\setlength{\tabcolsep}{1.8pt}
\begin{center}
\caption[]{Statistical results for case 5: $N = 1$, \quad $L = K$,\quad$M = 20$.}	\label{tab:statsc5}
		{ \small \begin{tabular}{ c | c | c | c | c | c | c | c | c | c | c | c | c | c}
			\hline \hline
			\textbf{N} & \textbf{K} & \textbf{M} & \textbf{L} & \textbf{T} & \textbf{$K_T$} & \textbf{A} & \textbf{Le} & \textbf{\%SIA} & $\mu$ & $\kappa$ & $\sigma A$ & $\sigma L$ & $\sigma \%SIA$ \\
			\hline \hline
			1 & 0.5 & 20 & 0.5 & 20 & 1 & 31.67 & 2 & 0.0068 & 0.05 & 0.5 & 51.596 & 0.870 & 0.0122 \\ \hline
			1 & 1 & 20 & 1 & 40 & 2 & 197.17 & 3 & 0.0591 & 0.05 & 0.5 & 182.155 & 1.615 & 0.0600 \\ \hline
			1 & 1.5 & 20 & 1.5 & 60 & 3 & 119.84 & 4 & 0.0416 & 0.05 & 0.5 & 124.516 & 2.668 & 0.0446 \\ \hline
			1 & 2 & 20 & 2 & 80 & 4 & 59.44 & 5 & 0.0243 & 0.05 & 0.5 & 69.834 & 4.101 & 0.0281 \\ \hline
		\end{tabular}}
	\end{center}
	
\end{table}


\begin{table}[!ht]
	\begin{center}
	\caption[]{Sensitivity to initial conditions for case 1: $K = L$,\quad $\mu\rightarrow1$.}
\label{tab:phasec1}
		{ \small \begin{tabular}{ c | c | c | c | c | c | c | c | c | c}
			\hline \hline
			\textbf{N} & \textbf{K} & \textbf{M} & \textbf{L} & \textbf{T} & \textbf{$K_T$} & $\Delta H$ & $\mu$ & $\kappa$ & $\sigma$ \\
			\hline \hline
			5 & 0.8333 & 4 & 0.8333 & 20 & 1 & -0.0402 & 1.25 & 0.8333 & 0.0293\\ \hline
			5 & 1.6666 & 4 & 1.6666 & 40 & 2 & -0.0067 & 1.25 & 0.8333 & 0.0798\\ \hline
			5 & 2.5 & 4 & 2.5 & 60 & 3 & 0.0582 & 1.25 & 0.8333 & 0.1406\\ \hline
			5 & 3.3333 & 4 & 3.3333 & 80 & 4 & 0.1799 & 1.25 & 0.8333 & 0.1936\\ \hline
		\end{tabular}}
	\end{center}
\end{table}

\begin{table}[!ht]
	\begin{center}
	\caption[]{Sensitivity to initial conditions for case 2: $K = 1$,\quad $\mu\rightarrow1$.}
\label{tab:phasec2}
		{ \small \begin{tabular}{ c | c | c | c | c | c | c | c | c | c}
			\hline \hline
			\textbf{N} & \textbf{K} & \textbf{M} & \textbf{L} & \textbf{T} & \textbf{$K_T$} & $\Delta H$ & $\mu$ & $\kappa$ & $\sigma$ \\
			\hline \hline
			5 & 1 & 4 & 0 & 20 & 1 & -0.0390 & 1.25 & 1 & 0.0293\\ \hline
			5 & 1 & 4 & 5 & 40 & 2 & -0.0081 & 1.25 & 0.5 & 0.0915\\ \hline
			5 & 1 & 4 & 10 & 60 & 3 & 0.0933 & 1.25 & 0.33333 & 0.1781\\ \hline
			5 & 1 & 4 & 15 & 80 & 4 & 0.2368 & 1.25 & 0.25 & 0.2051\\ \hline
		\end{tabular}}
	\end{center}
\end{table}

\begin{table}[!ht]
	\begin{center}
	\caption[]{Sensitivity to initial conditions for case 3: $L = 1$,\quad $\mu\rightarrow1$.}
\label{tab:phasec3}
		{ \small \begin{tabular}{ c | c | c | c | c | c | c | c | c | c}
			\hline \hline
			\textbf{N} & \textbf{K} & \textbf{M} & \textbf{L} & \textbf{T} & \textbf{$K_T$} & $\Delta H$ & $\mu$ & $\kappa$ & $\sigma$ \\
			\hline \hline
			5 & 0.8 & 4 & 1 & 20 & 1 & -0.0386 & 1.25 & 0.8 & 0.0350\\ \hline
			5 & 1.8 & 4 & 1 & 40 & 2 & -0.0107& 1.25 & 0.9 & 0.0698\\ \hline
			5 & 2.8 & 4 & 1 & 60 & 3 & 0.0327 & 1.25 & 0.93333 & 0.1062\\ \hline
			5 & 3.8 & 4 & 1 & 80 & 4 & 0.0832& 1.25 & 0.95 & 0.1413\\ \hline
		\end{tabular}}
	\end{center}
\end{table}

\begin{table}[!ht]
	\begin{center}
	\caption[]{Sensitivity to initial conditions for case 4: $M = 1$, \quad $N = 20$,\quad$L = 0$,\quad  $K=$$K_T$.}
\label{tab:phasec4}
		{ \small \begin{tabular}{ c | c | c | c | c | c | c | c | c | c}
			\hline \hline
			\textbf{N} & \textbf{K} & \textbf{M} & \textbf{L} & \textbf{T} & \textbf{$K_T$} & $\Delta H$ & $\mu$ & $\kappa$ & $\sigma$ \\
			\hline \hline
			20 & 1 & 1 & 0 & 20 & 1 & -0.04536 & 20 & 1 &0.0231\\ \hline
			20 & 2 & 1 & 0 & 40 & 2 & -0.0111 & 20 & 1 & 0.0930\\ \hline
			20 & 3 & 1 & 0 & 60 & 3 & 0.0964 & 20 & 1 &0.1806\\ \hline
			20 & 4 & 1 & 0 & 80 & 4 &0.2381 & 20 & 1 & 0.2076\\ \hline
		\end{tabular}}
	\end{center}
\end{table}

\begin{table}[!ht]
	\begin{center}
	\caption[]{Sensitivity to initial conditions for case 5: $N = 1$, \quad $L = K$,\quad$M = 20$.}
\label{tab:phasec5}
		{ \small \begin{tabular}{ c | c | c | c | c | c | c | c | c | c}
			\hline \hline
			\textbf{N} & \textbf{K} & \textbf{M} & \textbf{L} & \textbf{T} & \textbf{$K_T$} & $\Delta H$ & $\mu$ & $\kappa$ & $\sigma$ \\
			\hline \hline
			1 & 0.5 & 20 & 0.5 & 20 & 1 & -0.0334 & 0.05 & 0.5 &0.0329 \\ \hline
			1 & 1 & 20 & 1 & 40 & 2 & -0.0059 & 0.05 & 0.5 &0.0632 \\ \hline
			1 & 1.5 & 20 & 1.5 & 60 & 3 & 0.0241 & 0.05 & 0.5 &0.1027 \\ \hline
			1 & 2 & 20 & 2 & 80 & 4 & 0.0606 & 0.05 & 0.5 & 0.1358 \\ \hline
		\end{tabular}}
	\end{center}
\end{table}

\end{document}